\documentclass[english]{article}
\usepackage[T1]{fontenc}
\usepackage[latin1]{inputenc}
\usepackage{babel}
\usepackage{pstricks}
\usepackage{epsfig}
\usepackage{cite}
\usepackage{amsfonts,amsbsy,amsmath}
\usepackage{amssymb}
\usepackage{subfigure}
\newcommand{\be}{\begin{equation}}
\newcommand{\ee}{\end{equation}}
\newcommand{\bea}{\begin{eqnarray}}
\newcommand{\eea}{\end{eqnarray}}

\newcommand{\Dsla}{{\mbox{\, $\!\!\not\!\!D$}}}
\newcommand{\Dhat}{\hat{D}}
\newcommand{\Dbar}{\bar{D}}

\newcommand{\Famunu}{{\bf F}^a_{\mu \nu}(x)}

\makeatletter

\providecommand{\LyX}{L\kern-.1667em\lower.25em\hbox{Y}\kern-.125emX\@}

\makeatother
\begin{document}
\vskip -4cm

\title{Adjoint modes as probes of gauge field structure}

\author{Antonio González-Arroyo and Robert Kirchner}

\maketitle

\rput(0,7){FTUAM-05-7}
\rput(0.65,6.5){IFT-UAM/CSIC-05-22}

\begin{center}
{Depto. de Física Teórica C-XI \\
and \\
Instituto de Física Teórica UAM-CSIC\\
Universidad Autónoma de Madrid \\ Cantoblanco
Madrid 28049 SPAIN }
\end{center}

\begin{abstract}
{\noindent 
We show how zero-modes and quasi-zero-modes of the Dirac operator
in the adjoint representation can be used to construct an estimate
of the  action density distribution of a pure gauge field theory,
which is less sensitive to the ultraviolet fluctuations of the 
field. This can be used to trace the topological structures present
in the vacuum.  The construction relies on the special properties 
satisfied by the supersymmetric zero-modes.}
\end{abstract}

\section{Introduction}

The dynamics of gauge fields is expected to encompass some of the
most fascinating  phenomena, as confinement and chiral
symmetry breaking, whose   non-perturbative character 
makes them difficult to understand and describe. Lattice Gauge Theory has
opened the way to a first principles calculational scheme of all the
non-perturbative properties of the theory. In this way considerable
evidence has accumulated over the years that  Quantum Chromodynamics (QCD)
does indeed possess this behavior. It is, nonetheless, important to know if it
is possible to have a conceptually simpler description of these phenomena in terms of
a restricted set of relevant degrees of freedom. 

With the discovery of the instanton~\cite{bpst} and the subsequent qualitative 
solution of the $U_A(1)$ problem~\cite{tHooft76a}, giving the 
$\eta^{\prime}$ its 
mass by its coupling to the $U_A(1)$ current anomaly, it became clear that 
topology plays an important role in the low energy behavior  of QCD.
A longstanding  question, which is still open, is whether the mechanism
responsible for the  spontaneous breakdown of chiral 
symmetry (S$\chi$SB) does involve topological configurations, with the 
instanton as fundamental structure. This point is connected with the 
more 
general one about the validity and usefulness of the semiclassical approach 
to describe this and other non-perturbative aspects of QCD. In this respect a
semiclassical picture of the vacuum was developed, the Instanton Liquid 
Model \cite{ILM_1, ILM_2, Shuryak96}, which has met a certain degree of 
success in
explaining some quantities. This picture  provides a popular 
mechanism for the spontaneous breakdown of chiral symmetry, through 
the Banks-Casher relation \cite{BANCASH}, which relates the  chiral condensate
with the non-zero density of low-lying modes of the Dirac operator. The
latter  have its origin in the individual zero modes which  the Atiyah
Singer index theorem attributes to each instanton. In a dilute situation, 
with small overlap between neighboring instantons, these modes appear as
quasi-zero modes contributing a finite density~\cite{DP}. 

Although the mechanism is fairly appealing, there are several criticisms
to the overall picture. 
For example, it has been suggested that the ILM as main mechanism for S$\chi$SB is 
inconsistent with the expansion in the number of colors $N$. 
The argument is that at large $N$ the instanton weight in the action 
is exponentially suppressed, while effects from quantum fluctuations decay 
according to $1/N$.  This implies that at large $N$ instantons are not likely 
to play a role in the breakdown of chiral symmetry, since their effect is 
washed out by quantum fluctuations. Therefore, according to this argument,
either  chiral symmetry is broken differently for large $N$ than for $N=3$, 
or S$\chi$SB will not be caused by instantons in QCD. 
Another point of debate   is the diluteness of the instanton vacuum. 
Other descriptions  based  on a more dense multi-instanton
setting~\cite{AGAPM95_01}, might still inherit the  same mechanism for 
S$\chi$SB~\cite{AGAYS}. 

All these questions  about the general role of topology and semiclassical
methods in QCD should be resolved within the framework of lattice  gauge theory.
For this purpose different methods have been developed to extract the global 
and the local topology of lattice gauge fields in thermal equilibrium.
The main problem one encounters in examining Monte Carlo generated lattice 
gauge field 
configurations, is that they are very rough, with sizable fluctuations at the 
scale of the lattice spacing. This {\em noise} dominates over the long
wavelength signal one is willing to investigate. The roughness is to be
expected, and is a reflection of the ultraviolet divergences of quantum field
theory. Indeed, it is well known that even for a free field theory,
continuous fields have zero-measure. In Quantum Mechanics the divergence is 
milder, but still differentiable paths have zero-measure. Thus, 
 the rough {\em aura }
of smooth gauge fields is essential in contributing sizably to the path
integral. This, sometimes ignored fact, tells us that in the semiclassical 
approximation, smooth configurations act as labels to denote the finite
probability regions centered around them. 

A frequently used method to solve the aforementioned roughness problem 
is the so called {\em cooling} method~\cite{COOLING,COOLING_2}. 
Under this generic term a number 
of different procedures have been developed, which  have in common that they 
produce a sequence of increasingly smooth configurations by locally
minimizing the action. Ideally, the method should produce the smooth
{\em labelling} configuration associated to the rough initial one.
However, if cooling is applied indefinitely one would reach a relative  minimum 
of the action, in which  most of the local information would  be lost. 
Thus, beyond being useful in determining global topological quantities as 
the topological susceptibility~\cite{SUSZCOOL}, its validity to determine 
the local topology of the field, has been heavily and severely 
criticized. Nevertheless, we think that  cooling is indeed a useful method 
when used judiciously and in certain situations. 
The main issue has to do with a hierarchy in the variations of the action
within the space of gauge field deformations. We know that variations 
associated to rough fields produce large changes in the action.
On the other hand, there is sometimes a subclass of  smooth gauge fields
configurations associated to ``almost flat'' directions. 
This is indeed the case  in a dilute instanton gas situation, for example.
This hierarchy then would translate into an equivalent hierarchy of 
``cooling time'' scales. 
Thus, when cooling is applied, movement in the steeper directions  happens
relatively fast, while that along the ``almost flat'' directions is much 
slower. In addition, since cooling is a local algorithm, the evolution of 
large momenta components starts earlier. 
In any case, the usefulness  of the semiclassical approach depends also on 
the observable being studied. As in every reduced degrees of freedom approach,
a la Ginzburg-Landau, the projected degrees of freedom have to capture the 
essence of the physics being studied. 
Accordingly, if physical results have to be extracted from cooling, they 
should be independent  of the number of cooling steps applied within a
certain window, or change in a prescribed and computable fashion. 

A different approach to disentangle the local topological structure of the
lattice field, is commonly referred to as smoothing. It is a local procedure
which substitutes a link with a weighted average of link paths.
The prototype of smearing is APE-smearing \cite{APE}, but as in the case of 
cooling there is a number of algorithms in use, which however all share the 
common feature of averaging in some sense the gauge field locally. As in the 
case of cooling, smearing has successfully been used in determining the 
topological susceptibility \cite{ALLES93,KIRCHNER97,hasenfratz98,degrand98}, 
but has been the object
of criticism when local topological questions are concerned.
It is frequently stated that from a conceptual point of view smoothing is 
preferable over cooling, since the continuum limit of the gauge field is not 
changed in the process. 
However, as in the case of cooling, an indefinite number of steps would end up
washing out all local structures. Therefore, similar considerations as before are
necessary.  

A more recent proposal to  study vacuum topology uses fermionic degrees of
freedom. Introduced some time ago \cite{FERMION}, fermionic methods rely on 
their  response to the background gauge field. They suffer 
from the traditional problems associated to the breaking of chiral symmetry 
on the lattice. With the discovery that lattice Dirac operators which satisfy the 
Ginsparg-Wilson relation, as in the domain Wall formulation \cite{Kaplan92a,
Shamir,ShamFur} or 
the overlap operator \cite{Neuberger98a,Neuberger98b}, possess a lattice 
remnant of chiral symmetry at finite lattice spacing and satisfy exact index 
theorems, fermionic methods have become fully applicable. 
In various works \cite{DeGrand01,DeGrand02,Gattringer01,Edwards02,Blum02,
Horvath02b,Teper02,Horvath03,Horvath04,Horvath05} these methods have been
used to investigate the question of the local topological structure of the
QCD vacuum.
It is argued that an instanton dominated vacuum will lead to a spectrum of 
low-lying eigenmodes of the Dirac operator, which originates from 
the mixing of the zero-modes each instanton would contribute if it were 
isolated. The local chirality of the near zero-modes is then used as a 
measure of the topological origin of the mode.
It has been found \cite{DeGrand01,DeGrand02, Hip02,Edwards02,Blum02} that the 
local chiral structure of the low-lying 
modes of the Dirac operator does not exclude the ILM as a microscopically 
accurate picture of the vacuum. However, it is still  a matter of debate whether the 
structure of local chirality found in the low-lying modes does indeed have a
topological origin closely related to instantons or follows from other 
structures~\cite{Horvath02b,Gattringer02,Horvath02c,Horvath03,Horvath04,Horvath05}.

In the present work we want to propose a new method to investigate the local 
topological structure of the gauge fields, which is a hybrid of the 
above.
It adheres to the idea of using fermionic quasi-zero-modes to investigate the 
underlying local topology of the configurations. However, in our proposal 
we use the  Dirac operator in the adjoint representation. 
One of the advantages of using adjoint zero-modes instead of fundamental ones is
that some of these modes do exactly mimic the structure of the 
action density for classical solutions of the equations of motion. 
These particular modes are the supersymmetric partners of the corresponding 
gauge fields. They possess  a  peculiar reality property which allows 
to distinguish them from other modes. Thus, for gauge field configurations 
which are classical solutions of the equations of motion, we can construct 
two functions $S_{\pm}$ which reproduce the shape of the self-dual and 
anti-self-dual parts (respectively) of the action density up to an overall
normalization.  Following the same reasoning as in \cite{Horvath02a,Horvath02b},
for a smooth background gauge field which  is ``almost'' self-dual (as for
example a configuration consisting of dilute (anti)instantons) we expect to 
find an ``almost'' supersymmetric solution from which we can reconstruct the
form of the underlying gauge field.

The main usefulness of this idea arises in the presence of quantum
fluctuations. The quantities  $S_{\pm}$ are less sensitive to the 
ultraviolet modes than the action density itself. This follows from the 
non-local (extended) nature of these observables. Thus, as we will see,
it is still possible to reconstruct the long wavelength features of the 
configuration and its topological structure in the presence of quantum 
fluctuations of limited size. 
This paper is devoted to presenting the observables, the numerical 
method and to displaying its behavior for smooth configurations,
as well as in the presence of controlled quantum fluctuations.
A detailed analysis of  Monte Carlo generated
configurations is  deferred to later works. For that, it is
desirable to use a chiral invariant lattice Dirac operator, as the overlap. 
In the present paper we have used Wilson fermions and our results indicate 
that even in the worst cases studied, the new observables  seem to extract 
the local topology of the configuration  after a very small (1-3) number
of cooling steps are applied.

The paper is organized as follows: In Section 2 we review the main continuum
formulas concerning the  low-lying eigenstates of the massless adjoint Dirac
operator (adjoint modes), on which our numerical method is based. 
First, we discuss the situation for solutions of the
classical equations of motion. We identify the supersymmetric zero-modes and 
discuss their properties. We then study the behavior of the observables for 
configurations in the vicinity of a solution.
In Section 3 we discuss the lattice setup, i.e.
the operators, the algorithms and the configurations used throughout our 
work. We also introduce the numerical method which we use to project onto 
the supersymmetric or the ``almost'' supersymmetric solution of the Dirac 
operator in a given background field. 
In Section 4 we apply the method to specific configurations in order  
to explore its ability to reconstruct the form of the underlying gauge field.
We start by considering (anti)self-dual configurations and we show how our
numerical method is capable of isolating the supersymmetric solution, and how
we can, with a high degree of accuracy,  reconstruct the shape of the action 
density from these solutions. 
We then proceed with more complex smooth fields such as the non self-dual case
of an instanton anti-instanton pair. 
Finally we introduce quantum fluctuations. This we do by 
applying a given number of {\em heating} Monte Carlo steps to an initial
one-instanton configuration. For large $\beta$ and  a small number of 
cooling
steps we have control that the underlying topological structure of the 
configuration is unchanged and we can check the ability of our observables to 
reconstruct it. We also present results for smaller values of $\beta$. Here, 
the quantum fluctuations can change the long wavelength structure of the 
configuration, so our method is then used in 
conjunction with cooling and smearing. Results are promising, but we believe
a more accurate  analysis has to be done employing a chiral invariant lattice 
Dirac operator. Finally, in Section 6 we will summarize our conclusions.

\section{Adjoint modes of the Dirac Operator}
\label{continuum}
In this section we will shortly revise the main formulas and  explain the
philosophy of our method in the continuum.
We will be using the notation collected in the Appendix. 
Let us consider  smooth  SU(N) Yang-Mills field 
configurations on 4-dimensional euclidean space-time. Although the topology of 
space-time is not essential for the construction, in order to explain 
the numerical applications on the lattice, we will take space-time to 
be given by a 4-dimensional  torus.
Gauge fields are then connections on a (SU(N)) bundle over the torus; 
the latter  are classified into topologically inequivalent classes determined 
by the twist sectors  $n_{\mu \nu}$    and the instanton number 
$Q={ (N-1) \over 4N} n_{\mu \nu} \tilde{n}_{\mu \nu} + q$, 
where $q$ is an integer\cite{THOOFT,BAAL82}~\footnote{In the mathematical 
literature for SU(2) these are associated to
Stiefel-Whitney and Chern classes respectively. The reader is referred to 
Ref.~\cite{DK} for a mathematically  rigorous introduction to the subject.}. 
It is customary to split the antisymmetric integer-valued twist tensor 
$n_{\mu \nu}$ into its spatial ($m_i={1 \over 2}\epsilon_{ i j k} n_{j k}$),
and temporal ($k_i= n_{0 i}$) components. In this form the instanton number
reads
\be
\label{instnum}
Q=q+ \left( \frac{N-1}{N} \right) \vec{k} \cdot \vec{m}, \ \ \ \ q  \in  Z 
\ee
Notice that the instanton 
number is not necessarily an integer since  ${1 \over 4} n_{\mu \nu}
\tilde{n}_{\mu \nu} = \vec{k}\cdot \vec{m}$ need not be a multiple 
of $N$. 

We will now consider spinor fields $\Psi(x)$ transforming  in the adjoint representation 
of the gauge group. One can  study the spectrum of the Dirac operator in the 
background of the gauge field configuration:
\be
\label{dirac}
\Dsla^{\tiny A} \Psi^{(\lambda)} = i \lambda \Psi^{(\lambda)}
\ee
The  eigenvalues $i \lambda$ are (at least)
twice degenerate as a consequence of (euclidean space) charge conjugation 
symmetry, which  follows from the reality of the adjoint covariant 
derivative. Thus, for any solution $\Psi^{(\lambda)}$ of Eq.~\ref{dirac} we 
can construct a  new one as follows:
\be
\label{symadjoint}
\Psi_c^{(\lambda)} \longrightarrow \gamma_5 C  \Psi^{(\lambda)\ *} 
\ee
where $*$ denotes complex conjugation. This can be easily proven 
 by complex conjugating Eq.~\ref{dirac}. 
The new solution is orthogonal to the previous one (this 
follows from the antisymmetry of $C$).  

The positive and negative values of $\lambda$ are exactly 
mapped onto each other by multiplying the eigenvector by $\gamma_5$.
In addition, there might be zero-modes ($\lambda=0$).  The Atiyah-Singer 
index theorem  expresses the index of
the Dirac operator (number of zero-modes of positive chirality minus the 
number of zero modes of negative chirality) in terms of the instanton number:
\be
\mbox{index} \Dsla^{\tiny A}=  2N\ Q   
\ee
Notice that although $Q$ need not be an integer, the index is always an 
even number. As a matter of fact, in the presence of non-zero twist
($\vec{k},\vec{m}\ne 0$), the gauge group is $SU(N)/Z(N)$ and the
topological charge should be computed in the adjoint representation, which
a faithful representation of this group. Then, the topological charge
becomes precisely equal to the integer $2NQ$.
It is easy to understand the evenness as due to the 
two-fold degeneracy mentioned previously, which affects zero-modes of  
each chirality. 

The positive and negative chirality 
zero-modes ($\psi_\pm$) 
   satisfy the following Weyl equations:
\begin{eqnarray}
\label{PCWE}
 \Dbar^{\tiny A} \psi_+ = 0 \\
   \label{NCWE}
       \Dhat^{\tiny A} \psi_- = 0
           \end{eqnarray}
           respectively.
Contrary to the situation for zero-modes of the Dirac operator in the 
fundamental representation, the shape of  adjoint zero-modes is in some 
cases directly expressible in terms of the action density of the gauge 
field. In particular, suppose that $A_\mu$ is a gauge field configuration
which is a solution of the euclidean equations of  motion (not necessarily
self-dual or anti-self-dual), then the following four-spinor:
\be
\label{mainzm}
\Psi^a(V,x)=  \frac{1}{8}\, \Famunu [\gamma_\mu, \gamma_\nu] V \\
\ee
is an adjoint zero-mode for any constant four-spinor $V$ 
\cite{ZEROMODES,ZEROMODES2 }. This 
provides, if non-zero,   two linearly independent  positive chirality $\Psi^a_+(x)$
and two negative  chirality $\Psi^a_-(x)$ zero modes. The gauge invariant 
densities $|\Psi^a_\pm(x)|^2$ are equal (with the normalization given in 
Eq.~(\ref{mainzm}) and unitary $V$) to the self-dual/anti-self-dual
parts of the action density respectively.  This is the crucial result that we are
using in what follows. Since, the spinor field Eq.~\ref{mainzm} can be
obtained by applying a supersymmetry transformation to the gauge fields, we
will refer to it by the term  ``supersymmetric zero-mode''.   

There is an interesting property satisfied by  $\Psi^a_\pm(V,x)$ which
will allow us to single it out when there are several zero-modes present.
To explain  it,  it is better to take the chiral representation for the Dirac
matrices given in appendix A. Choosing $V^\dagger=\begin{pmatrix} 1 & 0 & 0
& 0 \end{pmatrix}$ one
gets an adjoint zero-mode of positive chirality such that the real part of
the first spinor component of  $\Re(\Psi^{a +}(V,x))$ vanishes in every space-time
point, for any value of the color index and all components in color-space.
Explicitly this positive chirality solution takes the form:
\begin{eqnarray}
\label{zeromodes}
\Psi^a_+= i \left(
\begin{array}{c}
\frac{(B^a_3+E^a_3)}{2}\\ 
\frac{(B^a_1+E^a_1)}{2}-i\frac{(B^a_2+E^a_2)}{2}\\ 
0 \\ 
0 
\end{array} \right ).
\end{eqnarray}
For $V^\dagger=(0 \ \ 1 \ \ 0 \ \ 0)$ the positive chirality solution 
$(\Psi^a_+)_c=\gamma_5C(\Psi^a_+)^*$ is found. In case of a self-dual 
gauge field, electric and magnetic field components coincide: 
$\vec{B^a}=\vec{E^a}$. 
The reality property of this zero-mode  is non-trivial: for a generic
solution it cannot be 
implemented by simply performing a 
gauge transformation or taking an appropriate choice of $V$.
The negative chirality solutions can be worked out analogously.
and the same conclusion follows. 

We have investigated whether the reality  property 
is shared by other zero-modes. In the simplest case of a single SU(2) instanton we 
have 4 positive chirality adjoint zero-modes, grouped in two charge-conjugate 
pairs. One of the pairs is the one corresponding  to Eq.~(\ref{mainzm}), and
the other does not satisfy the property. We have also investigated the
general ADHM~\cite{ADHM} case, which expresses adjoint zero-modes in terms
of the ADHM data.  Except in exceptional limits, this property only takes 
place for this special zero-mode. 

After this preamble we will now explain the philosophy of our method.
Consider the case of a general gauge field configuration,   not 
necessarily an exact solution of the euclidean classical equations of motion. 
Our strategy is to define a preferred spinor field in each chirality 
sector (a preferred section). 
The corresponding densities give rise to  two positive definite functions 
of space-time $S_\pm$. Their structure will trace that of the gauge field
from which they are constructed. We will exploit the previous considerations  
by requiring that for gauge fields which are  solutions of the classical 
equations of motion these quantities $S_\pm$ will coincide with the 
supersymmetric mode densities mentioned previously. Thus, they will be 
directly  proportional to the self-dual or anti-self-dual parts of
the action density (for left and right chiralities respectively). However, 
the main advantage of our observables compared to the action densities 
themselves appears when considering  configurations containing quantum
fluctuations. Our observables will be  better behaved in the ultraviolet.

There can be several ways to make the choice of section in order to 
implement the program.   It is at present unclear to us if there is an optimal 
choice. A particular simple and elegant possibility is to take as preferred 
section the eigenfunction of lowest eigenvalue (ground state) of the  positive-hermitian
operators $O_\pm$, given by the restrictions of  $-D\bar{D}$ 
(for left spinors) and $-\bar{D} D$ (for right spinors) to the space of states satisfying 
the reality condition. These operators  can be rewritten as
 real symmetric operators  acting on the space of real three-component
functions as follows:
\be
(O_\pm)^{a b}_{i j } = \Delta^{a b} \delta_{i j } \pm 4 F^{\pm\, c}_k
\epsilon_{ i  k j} f_{ a c b }
\ee
where $\Delta^{a b}$ is the covariant laplacian in the adjoint
representation and $F^{\pm\, c}_k= \frac{1}{2} (E^c_k \pm B^c_k)$. The
indices $a,b,c$ represent color in the adjoint representation and $i,j,k$
 run from 1 to 3. 
 
 The considerations made earlier in this section show that for a classical 
 solution of the euclidean equations of motion, $O^{(0)}_\pm$ has 
 a non-degenerate  ground state
with vanishing eigenvalue, and the densities $S_\pm$ are the self-dual and
anti-self-dual action densities. If we now perturb the gauge field configuration,
it is clear that the non-degeneracy of the ground state is preserved for
small enough deformations. One can use perturbation theory to compute the 
structure of the ground state in this case.  The variation of the ground state
(and its density) does not agree with the variation of the action density. 
Instead, it is expressed in terms of the Green function of the operator 
$O^{(0)}_\pm$. This a non-local expression which suppresses the 
 small wavelength fluctuations with respect to the large wavelength ones. 
 This is the effect we were looking for. 

For technical  reasons, related to the availability of code and data,
in the numerical part of this paper 
we have chosen a definition of the densities which is slightly different 
from the one presented in the previous paragraph. Rather than imposing
reality and looking for the ground states of $O_\pm$, we have searched
within the space of low lying modes of the Dirac operator, those which 
best satisfy the reality condition.  This is a change of order of the 
basic ingredients which, as  we will see in the following sections,
also gives rise to the required behavior.

\section{Lattice Methods}

\subsection{Lattice Dirac operator}
\label{lattice}
In order to translate the above  procedure  to the lattice,  a 
suitable discretization of the adjoint  Dirac operator has to be adopted.
For this pilot study we have chosen  Wilson discretization in the 
adjoint representation:  
\begin{eqnarray} 
\label{lat_dirac}
(D_W)_{ab}^{\alpha\beta}(x,y) &=& 4r \delta_{ab} \delta_{xy} 
\delta^{\alpha\beta} \nonumber \\ 
&-& \frac{1}{2}\sum_{\mu=1}^{4} 
\left[ \delta_{y,x+\hat{\mu}}(r+\gamma_{\mu})^{\alpha \beta} V_{\mu,ab}(x) \right. \nonumber \\
&+&  \left. \delta_{y+\hat{\mu},x}(r-\gamma_{\mu})^{\alpha \beta} 
(V^T)_{\mu,ab}(y) \right], \\ \nonumber
\end{eqnarray}
where $V_{\mu,ab}(x)$ is the link in the adjoint representation of the gauge 
group (see the Appendix, Eq. (\ref{adj_link})).
\newline
The above operator has the following properties:
\begin{eqnarray*}
C^{-1}D_W C=D_W^{T}, \\
\gamma_5 D_W^{\dagger} \gamma_5 =D_W, 
\end{eqnarray*}
which implies that $(\gamma_5 D_W)$ is hermitian.

For our numerical work we will consider the eigenstates of the positive 
operator $(\gamma_5 D_W)^2$.
Because of the symmetry properties described by Eq. (\ref{symadjoint}) the 
eigenspaces  are (at least) two-fold degenerate.

Let us comment on the eigenvalue structure of $(\gamma_5 D_W)^2$ for 
self-dual or anti self-dual background fields. 
From the Atiyah-Singer index theorem we know that for a massless continuum 
Dirac operator in the adjoint representation of the gauge group, and in the 
background of a $SU(N)$ gauge field carrying charge $Q$, we will 
find $ \rm{Index}(D\!\!\!\!/)=2N|Q|$ zero-modes, with definite chirality.
On the lattice, using $(\gamma_5 D_W)^2$ as Dirac operator, this
translates to finding $2N|Q|$ eigenvectors with ``small'' eigenvalues and 
given chirality. 
These ``small'' eigenvalues are separated by a gap from a more densely 
populated band. 
In what follows we will refer to the modes with small eigenvalues in the 
above sense as approximate zero-modes.
\newline
Another important point concerns the chiral properties 
of  the Wilson-Dirac operator. The explicit breaking of chiral symmetry by 
this  operator might  pose a problem, since the 
properties of the eigenmodes of the Dirac operator we wish to study on the 
lattice depend very much on chirality.  The smaller the value of the
parameter r, the smaller is the breaking of naive chirality. However, this 
conflicts with the splitting given to the doublers. Our final choice has been 
to work at the small value $r=0.3$ of the Wilson parameter. 
As we will verify a posteriori this choice 
is small enough so that the properties we wish to study 
that depend on chirality are not too seriously affected. 
To check that there is no sizeable contamination of doublers in the low edge of the spectrum
that we study, 
we computed the  matrix elements in the corresponding eigenspace of the
Wilson operator, defined by:
\begin{eqnarray} 
\label{wilson}
(W)_{ab}^{\alpha\beta}(x,y) &=& 4 \delta_{ab} \delta_{xy} 
\delta^{\alpha\beta} \nonumber \\ 
&-& \frac{1}{2}\sum_{\mu=1}^{4} 
\left[ \delta_{y,x+\hat{\mu}}\delta^{\alpha\beta} V_{\mu,ab}(x) \right. \nonumber \\
&+&  \left. \delta_{y+\hat{\mu},x}\delta^{\alpha\beta}(V^T)_{\mu,ab}(y) \right], \\ \nonumber
\end{eqnarray}
This procedure has been used previously by our group and is known to work
well for smooth gauge field configurations.
For rough, Monte Carlo generated configurations  it is certainly desirable to 
use a discretization which is more respectful of chiral symmetry, 
such as the Neuberger operator \cite{Neuberger98a,Neuberger98b}.  
Nevertheless, for the sake of our paper we concluded that  the much less 
computationally costly Wilson-Dirac operator still allowed us to exemplify the validity
of our idea.

\subsection{Projecting onto the supersymmetric solution}
\label{projection}
In Section \ref{continuum} we have defined our procedure  in the continuum. 
Our goal is to construct two positive
functions  of space $S_{\pm}$  for each gauge field  configuration.
For the case that the gauge field is a solution of the Euclidean equations of 
motion $S_{\pm}$ they reproduce exactly the shape of the self-dual and anti-self-dual
parts of the action density. Deforming away from them, 
we have  argued that these quantities
are less sensitive to the ultraviolet modes of the gauge field than the 
action density itself. Thus,  in the presence of quantum fluctuations they
produce an estimate of the long-range topological structure of the
configurations.
\newline
To provide a lattice implementation, let us first consider the case 
in which the link variables are  a lattice discretization of a background 
gauge field which is a solution of the classical Euclidean equations of 
motion having  a non-trivial self-dual part. In this case we first  
have to look at the low-lying eigenstates of the   Wilson-Dirac operator. 
This space of  states should contain the discretized zero-modes. We then explore 
within this space to find the state that best projects onto the positive chirality
supersymmetric solution Eq.(\ref{zeromodes}) of the Weyl equation.
To single out this solution we make use of the
reality property of the cuadrispinor of Eq.(\ref{mainzm}) discussed in
Section \ref{continuum}.
On the lattice this is done by looking at the linear combination of approximate 
zero-modes that has minimal   imaginary part of the first spinor component 
for all colors and at each point of space-time.  We also require that the solution
is as chiral as possible. 
If we indicate by $\phi_i$  the elements of the set of linearly independent quasi-zero
modes, then our solution can be written as 
\begin {equation}
\label{numsol}
\Psi_+=\lambda_i \phi_i
\end{equation}   
The  coefficients $\lambda_i$ are chosen such that 
\begin{equation}
\label{cond}
c= \sum_{x,a}(\rm{Im} \Psi_+^{1,a}(x))^2 + \sum_{x,a,s=3,4}(\rm{Re} 
\Psi_+^{s,a}(x))^2 + (\rm{Im} \Psi_+^{s,a}(x))^2 
\end{equation}
is minimal. 

In the same way,  for a field having a non-trivial anti self-dual part, 
the negative chirality solution $\Psi_-$ corresponding to Eq.(\ref{zeromodes}) in the 
continuum, can be calculated by 
minimizing in Eq.(\ref{numsol}) the imaginary part of the third component 
and the first and second component (chirality).

In terms of the coefficients  $\lambda_i$ Eq.~(\ref{cond}) can be written as a quadratic form:
\begin{equation}
\label{M}
c= \left< v,Mv \right>, \ \ {\rm with } \ \ \ \ 
v=\left(
\begin{array}{c}
\rm{Re} \lambda_i\\   
\rm{Im} \lambda_i 
\end{array} \right ), \ \
M=\left(
\begin{array}{cc}
\rm{A} & \rm{B}\\   
\rm{B^T} & C
\end{array} \right ).
\end{equation}
The matrix M is $2n$ dimensional, where n is the number of eigenstates of the 
Dirac operator among which the minimization is carried out. 
It is straightforward to work out the detailed form of 
the sub-matrices A, B and C:
\begin{eqnarray*}
&&
A_{k,l}=\sum_{x,a}\rm{Im}\phi_k^1 \rm{Im}\phi_l^1 + \sum_{x,a,s=3,4} \rm{Im}\phi_k^s \rm{Im}\phi_l^s + \rm{Re}\phi_k^s \rm{Re}\phi_l^s , \nonumber \\&&
B_{k,l}=\sum_{x,a}\rm{Im}\phi_k^1 \rm{Re}\phi_l^1 + \sum_{x,a,s=3,4} \rm{Im}\phi_k^s \rm{Re}\phi_l^s - \rm{Re}\phi_k^s \rm{Im}\phi_l^s, \\&&
C_{k,l}=\sum_{x,a}\rm{Re}\phi_k^1 \rm{Re}\phi_l^1 + \sum_{x,a,s=3,4} \rm{Re}\phi_k^s \rm{Re}\phi_l^s + \rm{Im}\phi_k^s \rm{Im}\phi_l^s ,
\end{eqnarray*}
$M$ is also hermitian, positive, and for normalized vectors $v$ the minimal 
eigenvalue of $M$ is the minimum of expression (\ref{cond}). The corresponding 
eigenvector $v$ provides the real and imaginary part of the coefficients 
$\lambda_i$.~\footnote{The complete diagonalization of $M$ can be carried 
out by standard libraries.}
The lattice approximation $\Psi_{\pm}$ to the supersymmetric solution can be 
obtained by using  to Eq.(\ref{numsol}) with the $\lambda_i$ obtained in the
previous procedure. 
\newline
Then, $S_{\pm}(x)$ are then defined as:
\be
 S_{+}(x)  = \left< (\Psi_+)_L,(\Psi_+)_L \right >, \ \ \ \
 S_{-}(x)  = \left< (\Psi_-)_R,(\Psi_-)_R \right > .
\ee
where brackets denote scalar products in spinor and color space.
Note that since the reality property is very restrictive, finding the vector
that minimizes expression (\ref{cond}) in a larger space  containing all  approximate
zero modes,  will, up to numerical errors, not influence result. This 
is crucial, since one does not know how many zero-modes one should find  for a 
generic gauge field. This observation will   also play a role later 
in testing  the applicability and 
stability of the above described numerical procedure.

\subsection{Technicalities}
\label{technical}
Let us briefly mention some technical details of this work.
For the numerical part we used $SU(2)$ configurations throughout.
The self-dual gauge configurations have in all cases 
been obtained by standard cooling \cite{COOLING,COOLING_2}. 
\newline
Next to cooling we also use APE-smearing \cite{APE} to smoothen the 
configurations in some cases.
The $\rm{N}^{th}$-level APE-smeared link with smearing parameter $c$ is defined
as:
\begin{eqnarray}
{\overline U}^{(i)}_{\mu}(x) &=& (1-c) U^{(i-1)}_{\mu}(x) +
{c \over 6} 
\sum_{{\scriptstyle \alpha = \pm 1} \atop { \scriptstyle 
|\alpha| \not= \mu}}^{\pm 4}
U^{(i-1)}_{\alpha}(x) U^{(i-1)}_{\mu}(x+\hat{\alpha})
U^{(i-1)}_{\alpha}(x+\hat{\mu})^{\dag}, \nonumber
\end{eqnarray}
with ${\overline U}^{(i)}_{\mu}(x)$ projected back in $SU(2)$ after each 
smearing step. For small smearing parameter c this corresponds to a diffusion 
process with a diffusion radius:
\begin{eqnarray}
R_{s}=Nc. \nonumber
\end{eqnarray}
To introduce short range ultraviolet fluctuations 
the heat-bath algorithm of \cite{HEATBATH} was used.
The lattice action used is the usual Wilson action. 
The topological charge operator we use to 
identify the charge of smooth gauge field configurations is the field 
theoretic operator discussed in \cite{QFIELDTHEOR}.
The low-lying eigenstates and their respective 
eigenvalues of the operator $(\gamma_5D_W)$ are calculated with the 
conjugate gradient algorithm proposed by Kalkreuther and Simma \cite{SIMMA}, 
and with the implicitly restarted Arnoldi algorithm \cite{ARNOLDI}. 
For lattice sizes
$8^4$, $12^4$ and $12^3 * 24$ we computed about 10 to 20 eigenstates.

\section{Results}

In this section we will present the results obtained by applying 
the numerical procedure of projecting onto the supersymmetric modes  of 
the adjoint Dirac operator  described in the previous sections. 
We will consider certain SU(2) gauge configurations on the lattice in an 
increasing order of complexity.  
First we investigate the case of smooth (anti)self-dual configurations.
This will test the efficiency of our method to identify the supersymmetric 
zero-mode Eq.~(\ref{zeromodes}) within  the  subspace of low-lying eigenstates. 
The situation becomes more difficult 
the larger the value of the topological charge $|Q|$, since  
the dimensionality of the zero-mode manifold ($ 4 |Q|$) increases. 
\newline
After,  we will also apply the procedure to smooth
configurations which are not solutions of the classical equations
of motion, and in particular the case of an instanton
anti-instanton pair. Thus, the solution will not be an exact zero-mode
of the Dirac operator anymore. Finally, 
we will conclude by adding quantum fluctuations to the gauge field 
configurations. This is done by subjecting them to a given number of 
Monte Carlo updating steps. 

\subsection{Smooth self-dual Configurations}
\label{sd_conf}
We begin by  applying our procedure to obtain the supersymmetric zero-mode 
of the adjoint Dirac operator for a number of (anti)self-dual configurations 
with increasing values of the topological charge $|Q|$ (see  Table
\ref{conf_sum}). The smallest absolute value of topological charge we consider 
is $Q=-\frac{1}{2}$, which demands the use of twisted boundary 
conditions ($\left< m,k \right > \neq 0 $) as explained in section 2.
\begin{table}[ht]
\begin{center}
\begin{tabular}{| c | l | l | c | l | l | l | l | l | l |}
\hline
lattice: $L,T$  &
\multicolumn{1}{|c|}{$S/(8\pi^2)$} &
\multicolumn{1}{|c|}{$Q_L$} &
\multicolumn{1}{|c|}{$\rm{Twist_m}$}  & {$\rm{Twist_k}$} &              
\multicolumn{1}{|c|}{$r $} &              
\multicolumn{1}{|c|}{$\rm{ \# Eigenst.} $ }  
              \\
\hline
8,8   & 0.493 &-0.47& 1 1 1 & 1 1 1 & 0.3 & 10       \\
\hline
12,12 & 1.02 &-0.90& 0 0 0 & 0 0 0 & 0.3& 16    \\
\hline
12,12 & 1.51 & -1.41& 1 1 1 & 1 1 1 & 0.3& 16  \\
\hline
12,12 & 1.97 & 1.86& 1 0 1 & 0 1 0 & 0.3  & 20  \\
\hline
\end{tabular}
\parbox{12cm}{\caption{\label{conf_sum} \it
Configurations for which the reconstruction of the supersymmetric zero-mode  
was carried out. The values of  $Q_L$ were obtained with the field-theoretic
unsmeared topological charge operator. The second column gives is the total action
$S$ divided by
$8\pi^2$.}}
\end{center}
\end{table}

\begin{enumerate}
{\bf \item $Q=-\frac{1}{2}$, m=\{111\}, k=\{111\} }
\newline
\newline
In the case of an anti self-dual $Q=-\frac{1}{2}$ field there are only two 
zero-modes 
which correspond to the supersymmetric solution  Eq. (\ref{zeromodes}). 
We have numerically calculated the ten lowest-lying modes of the 
operator $(\gamma_5 D_W)^2$ in this background. The results are summarized 
in Table \ref{q=-0.5}. Notice that the eigenvalues are twice degenerate 
as follows from the  symmetry $\phi_i=\gamma_5 C^{-1} \phi^*_{i+1}$ (see
section 2). In the Table we also show the chirality ($ \left < \phi, \gamma_5 \phi \right >$) of each mode.  

\begin{table}[h]
\begin{center}
\begin{tabular}{| c | l |}
\hline
$\rm{Eigenvalue}$  &
\multicolumn{1}{|c|}{$\left < \phi, \gamma_5 \phi \right > $} 
 \\
\hline
 $\lambda_1$=2.44 * $10^{-3}$ & -0.998 \\
\hline
 $\lambda_3$=2.68 * $10^{-1}$ & 0.108    \\
\hline
 $\lambda_5$=2.68 * $10^{-1}$ & 0.108 \\
\hline
 $\lambda_7$=2.68 * $10^{-1}$ & 0.108 \\
\hline
 $\lambda_9$=2.93 * $10^{-1}$ & -0.106 \\
\hline
\end{tabular}
\end{center}
\begin{center}
\parbox{12cm}{\caption{\label{q=-0.5} \it
Lowest eigenvalues  of the operator $(\gamma_5 D_W)^2$, and the chirality of 
the corresponding modes, for the configuration with $Q=-\frac{1}{2}$.}}
\end{center}
\end{table}

There is a clear gap of two orders of magnitude between the first two
eigenvalues and the higher states. 
Accordingly, we can identify the modes pertaining
to $\lambda_{1,2}$ with the zero-modes. This is also clear from the
chirality of the modes. To confirm that their  structure
is the one given by Eq.(\ref{zeromodes}), we compare  the chiral density
\begin{equation} 
\left< \Psi_R,\Psi_R \right >
\propto_{a 
\rightarrow 0}
\rm{Tr}[F_{\mu\nu}F_{\mu\nu}] +O(a^2), 
\label{chiral_dens}
\end{equation}
of the lowest lying mode with the action density. The shapes are displayed 
in Figure \ref{fig_q=-0.5} for  a given plane, and show a nice 
match up to  an overall normalization. The latter might be  fixed
by the value of the topological charge alone.

\begin{figure}
\centerline {\epsfig {file=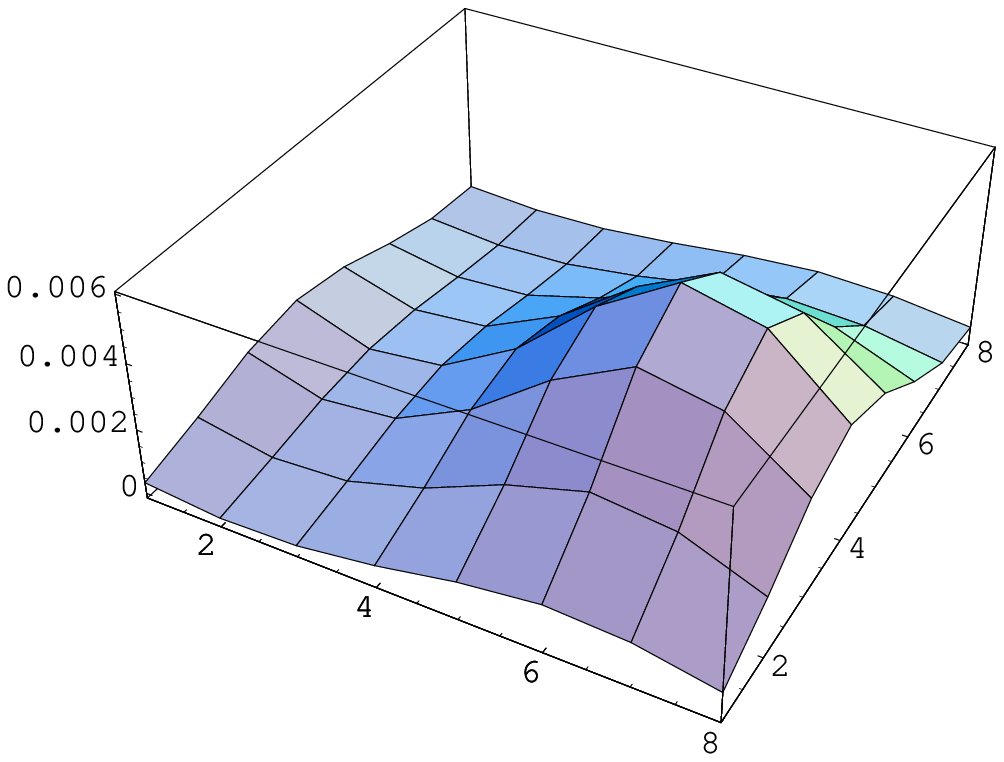 , 
width=6cm, height=6cm}
\epsfig {file=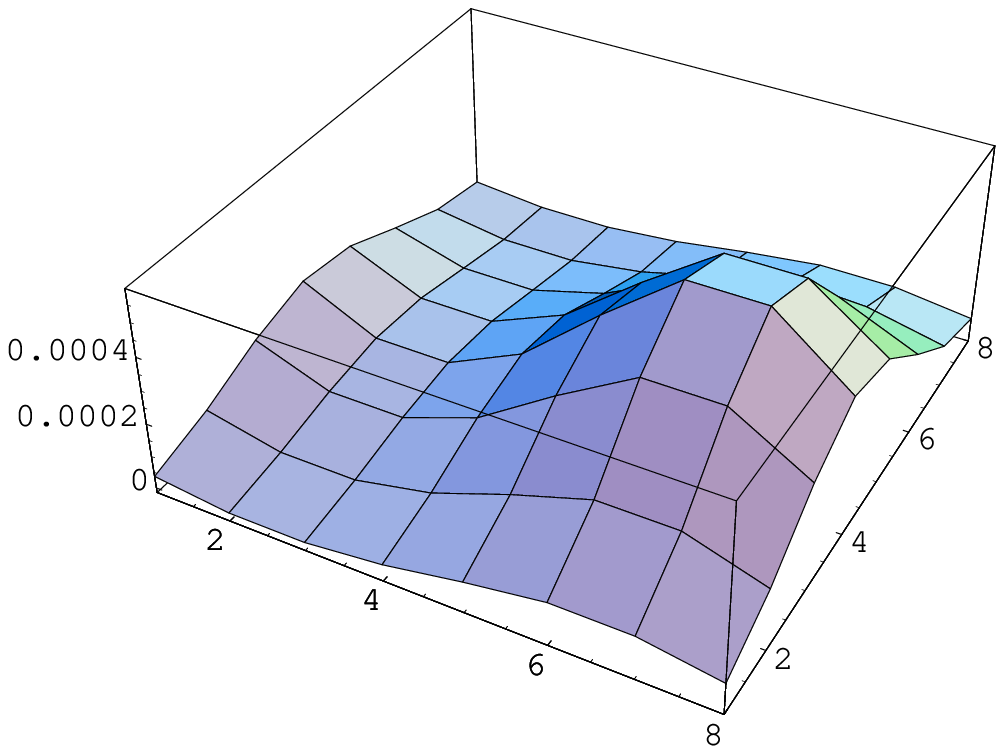 , 
width=6cm, height=6cm} }
\caption{ \it Action density $F^2$ (left) and chiral density 
$\left<\Psi_R, \Psi_R\right>$ (right) along a  t-z plane, 
for the configuration with $Q=-\frac{1}{2}$ .}
\label{fig_q=-0.5}
\end{figure}
It is important to test the stability of the method with respect to the 
number of eigenmodes considered, since ultimately one wants to apply 
the method to situations for which the  number of quasi-zero modes is not 
determined a priori. 
For that purpose, we have considered the full space of ten eigenmodes and applied 
the minimization condition  (\ref{cond}) to the  matrix $M$ (see
Section~\ref{projection}). The vector that achieves the minimum 
is indeed  dominated by the two 
low-lying modes, with the remaining modes  giving contributions two orders of
magnitude smaller. The density profile is only slightly changed.

{\bf \item $Q=-1$, m=k=\{000\} }
\newline
\newline
For an anti self-dual $Q=-1$ configuration the space of zero-modes is 
four dimensional. To find the zero-mode of Eq. (\ref{zeromodes})
we have calculated 
the sixteen lowest-lying eigenstates of $(\gamma_5 D_W)^2$ in this background.
The results are shown in 
Table \ref{Q=-1}. We find four 
small eigenvalues which are separated by one order of magnitude from 
the higher ones. 
The corresponding eigenvectors have definite chirality, and we identify these 
modes as the zero-modes predicted in the continuum by the Atiyah-Singer 
index theorem.
\begin{table}[h]
\begin{center}
\begin{tabular}{| c | l | l | l |}
\hline
$\rm{Eigenvalue}$  &
\multicolumn{1}{|c|}{$\left < \phi, \gamma_5 \phi \right > $} &
$\rm{Eigenvalue}$  &
\multicolumn{1}{|c|}{$\left < \phi, \gamma_5 \phi \right > $} 
 \\
\hline
 $\lambda_1$=7.79 * $10^{-5}$ & -0.997 & $\lambda_9$=5.11 * $10^{-2}$& 0.057 \\
\hline
 $\lambda_3$=7.40 * $10^{-3}$ & -0.976  & $\lambda_{11}$=5.34 * $10^{-2}$ & -0.057  \\
\hline
 $\lambda_5$=1.35 * $10^{-2}$ & 0.178 * $10^{-2}$ & $\lambda_{13}$=7.49 * $10^{-2}$ & 0.067 \\
\hline
 $\lambda_7$=2.25 * $10^{-2}$ & -0.164 * $10^{-2}$ & $\lambda_{15}$=8.41 * $10^{-2}$ & -0.075  \\
\hline
\end{tabular}
\end{center}
\begin{center}
\parbox{12cm}{\caption{\label{Q=-1} \it
Lowest eigenvalues of the operator $(\gamma_5 D_W)^2$ and chiralities of the 
respective eigenstates, for the configuration with $Q=-1$.}}
\end{center}
\end{table}   
In principle the solution (\ref{zeromodes}) that we are looking for is
a linear combination in the space of these zero-modes.
Applying the minimization within this four dimensional space, 
  we indeed find that the minimizing vector
$\Psi_-$ has the expected structure given by Eq.~(\ref{zeromodes}).
We exhibit this by plotting
the chiral density (\ref{chiral_dens}) in a fixed plane and comparing  it to the 
action density in the same plane (see Figure \ref{fig_q=-1}).
Including in the minimization all eigenstates that we have calculated, the same
linear combination is selected 
(new coefficients are suppressed by two orders of magnitude) as in the space
of zero-modes. Accordingly, also in the case of $Q=-1$ we
obtain stable results.  
\begin{figure}[ht]
\centerline {\epsfig {file=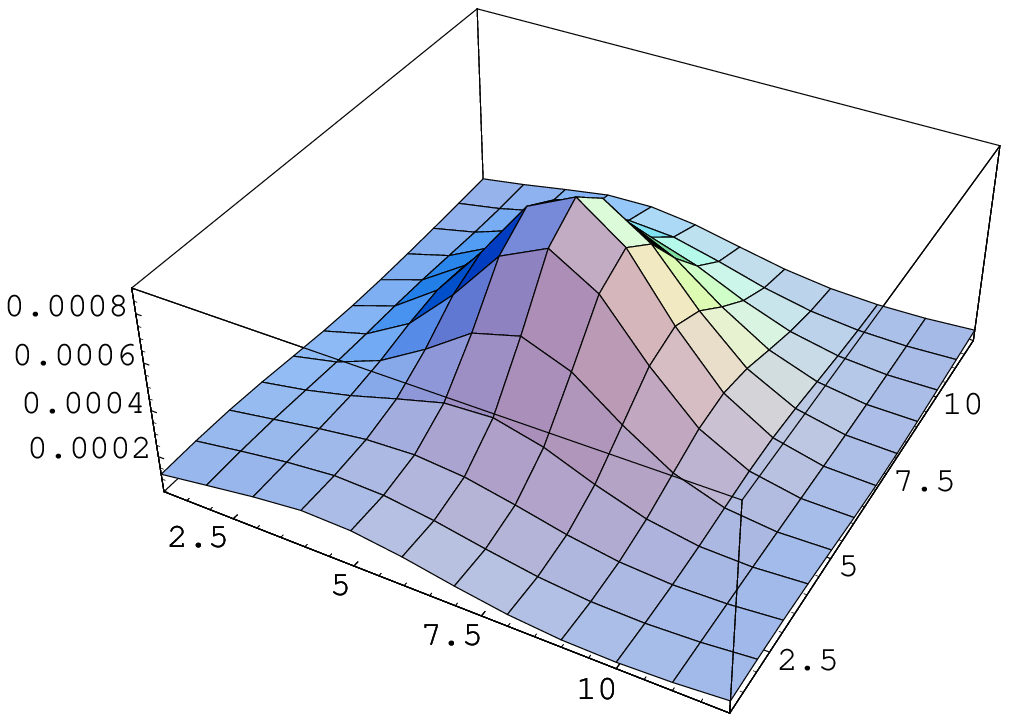, 
width=5cm, height=5cm}
\epsfig {file=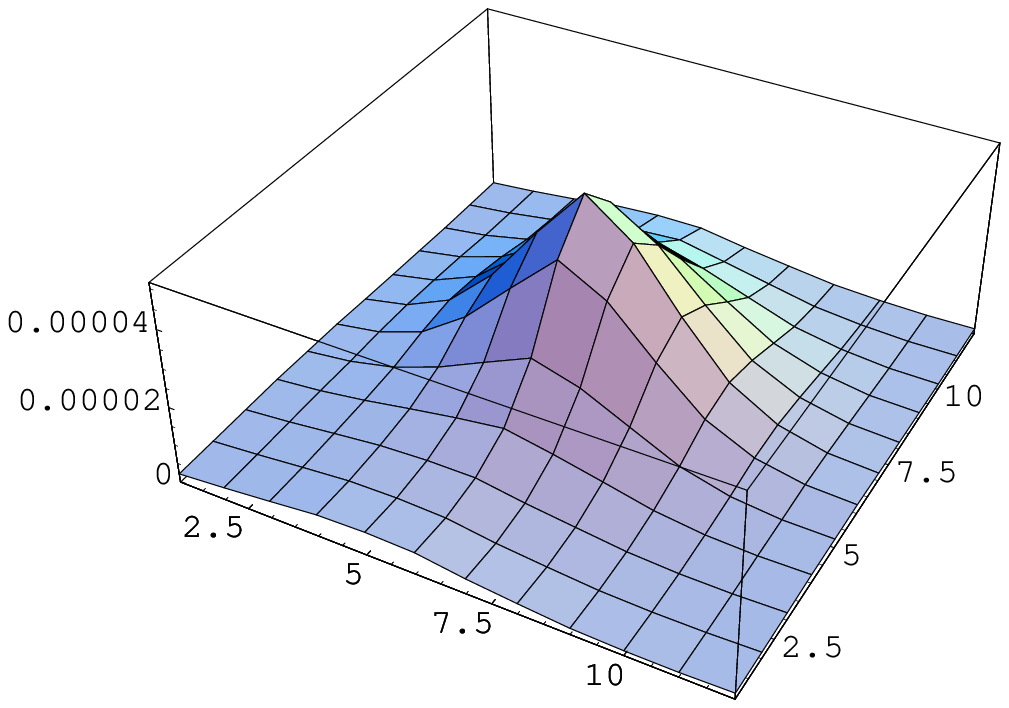, 
width=5cm, height=5cm} }
\caption{ \it Action (left) and chiral (right) densities for the $Q=-1$ 
configuration, for a given t-z plane.}
\label{fig_q=-1}
\end{figure}

{\bf \item $Q=-1.5$, m=\{111\}, k=\{111\} }
\newline
\newline
In the case of the $Q=-1.5$ configuration we find 6 low-lying modes that are 
divided by a gap of about two orders of magnitude from a band of higher 
eigenvalues (see Table \ref{Q=1.5}).
Again we find by minimizing expression (\ref{cond}) in the space of
these lowest lying modes that the resulting linear combination is of the 
form of Eq. (\ref{zeromodes})
(see Figure \ref{fig_q=1.5}). Including all higher eigenmodes in the 
minimization does not change the linear combination appreciably, and hence, 
also in this case the minimization is stable. 
\begin{table}[h]
\begin{center}
\begin{tabular}{| c | l | l | l |}
\hline
$\rm{Eigenvalue}$  &
\multicolumn{1}{|c|}{$\left < \phi, \gamma_5 \phi \right > $} &
$\rm{Eigenvalue}$  &
\multicolumn{1}{|c|}{$\left < \phi, \gamma_5 \phi \right > $} 
 \\
\hline
$\lambda_1$=9.00 * $10^{-4}$ & -0.997 & $\lambda_9$=1.16 * $10^{-1}$ & -0.085\\
\hline
$\lambda_3$=1.38 * $10^{-3}$ &-0.995  & $\lambda_{11}$=1.23 * $10^{-1}$& 0.086 \\
\hline
$\lambda_5$=3.25 * $10^{-3}$ & -0.990  &$\lambda_{13}$=1.32 * $10^{-1}$ & 0.077 \\
\hline
 $\lambda_7$=1.05 * $10^{-1}$ & 0.090 &$\lambda_{15}=1.36$ * $10^{-1}$ & -0.082\\
\hline
\end{tabular}
\end{center}
\begin{center}
\parbox{12cm}{\caption{\label{Q=1.5} \it
Lowest lying eigenvalues of the operator $(\gamma_5 D_W)^2$ and chiralities of 
the respective eigenstate, for the configuration with charge $Q=-\frac{3}{2}$.}}
\end{center}
\end{table}

\begin{figure}[ht]
\centerline {\epsfig {file=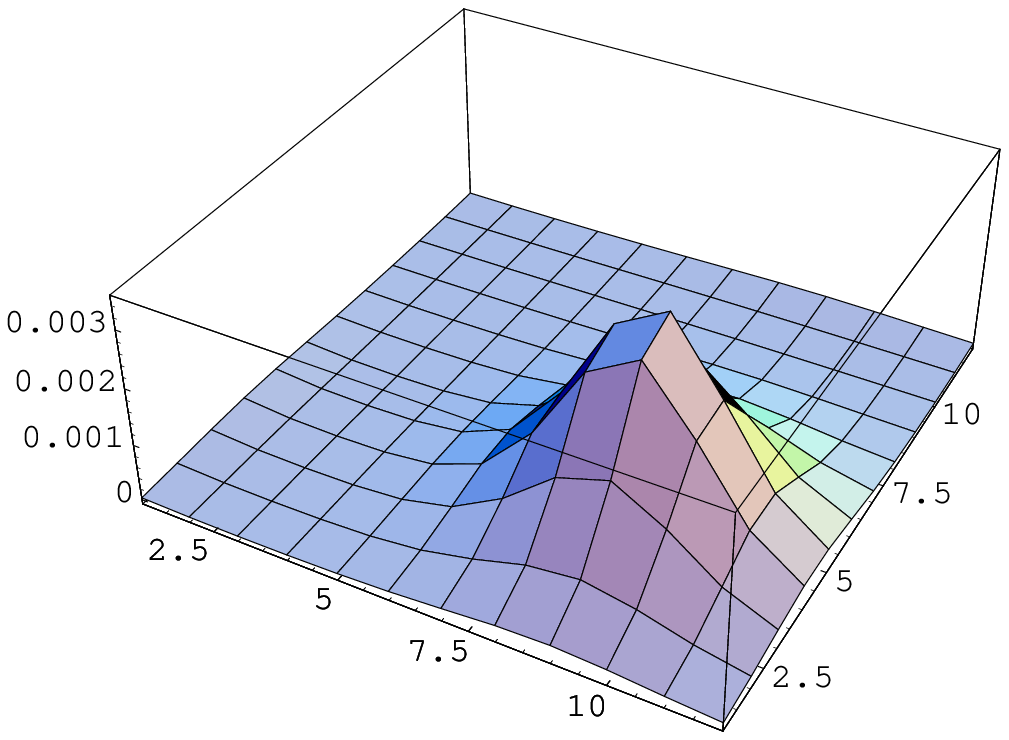, width=5cm, height=5cm}
\epsfig {file=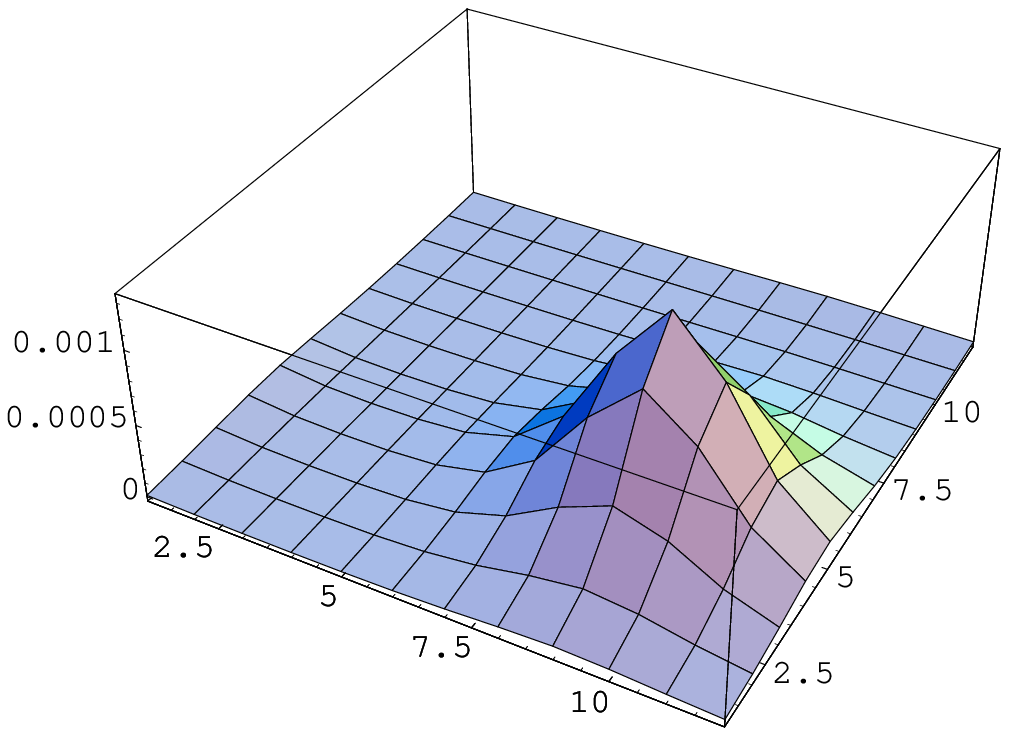, 
width=5cm, height=5cm} }
\caption{ \it Action (left) and chiral density (right) for the $Q=-1.5$ 
configuration; both in the t, z plane.}
\label{fig_q=1.5}
\end{figure}

{\bf \item $Q=2$, m=\{101\}, k=\{010\} }
\newline
\newline
The next case we consider in order to test our method is $Q=2$. 
We
find eight low-lying modes (see Table \ref{Q=2}) which are divided by an 
order of magnitude 
from the higher eigenvalues. As above, minimizing 
Eq.~(\ref{cond}) selects the wave function Eq.~(\ref{zeromodes}) among the 
eight lowest lying modes which we associate to the zero-modes of the continuum 
configuration (see Figure \ref{fig_q=2}). 
As in all other cases we observe stability of the 
projection onto the supersymmetric mode when the higher modes are included.
\begin{table}[h]
\begin{center}
\begin{tabular}{| c | l | l | l |}
\hline
$\rm{Eigenvalue}$  &
\multicolumn{1}{|c|}{$\left < \phi, \gamma_5 \phi \right > $} &
$\rm{Eigenvalue}$  &
\multicolumn{1}{|c|}{$\left < \phi, \gamma_5 \phi \right > $} 
 \\
\hline
 $\lambda_1$=6.28 * $10^{-4}$ & 0.997 & $\lambda_{11}$=8.87 * $10^{-2}$ &0.090  \\
\hline
 $\lambda_3$=1.20 * $10^{-3}$ & 0.995 & $\lambda_{13}$=1.02 * $10^{-2}$ &-0.085  \\
\hline
 $\lambda_5$=1.98 * $10^{-3}$ & 0.991 & $\lambda_{15}$=1.14 * $10^{-2}$ &0.084 \\
\hline
 $\lambda_7$=7.27 * $10^{-3}$ & 0.982 & $\lambda_{17}$=1.21 * $10^{-2}$ &-0.085 \\
\hline
 $\lambda_9$=7.50 * $10^{-2}$ & -0.094 & $\lambda_{19}$=1.30 * $10^{-2}$ &-0.080 \\
\hline
\end{tabular}
\end{center}
\begin{center}
\parbox{12cm}{\caption{\label{Q=2} \it
Lowest lying eigenvalues of the operator $(\gamma_5 D_W)^2$ and chiralities 
of the respective eigenstate, for the configuration with $Q=2$.}}
\end{center}
\end{table}

\begin{figure}[ht]
\centerline {\epsfig {file=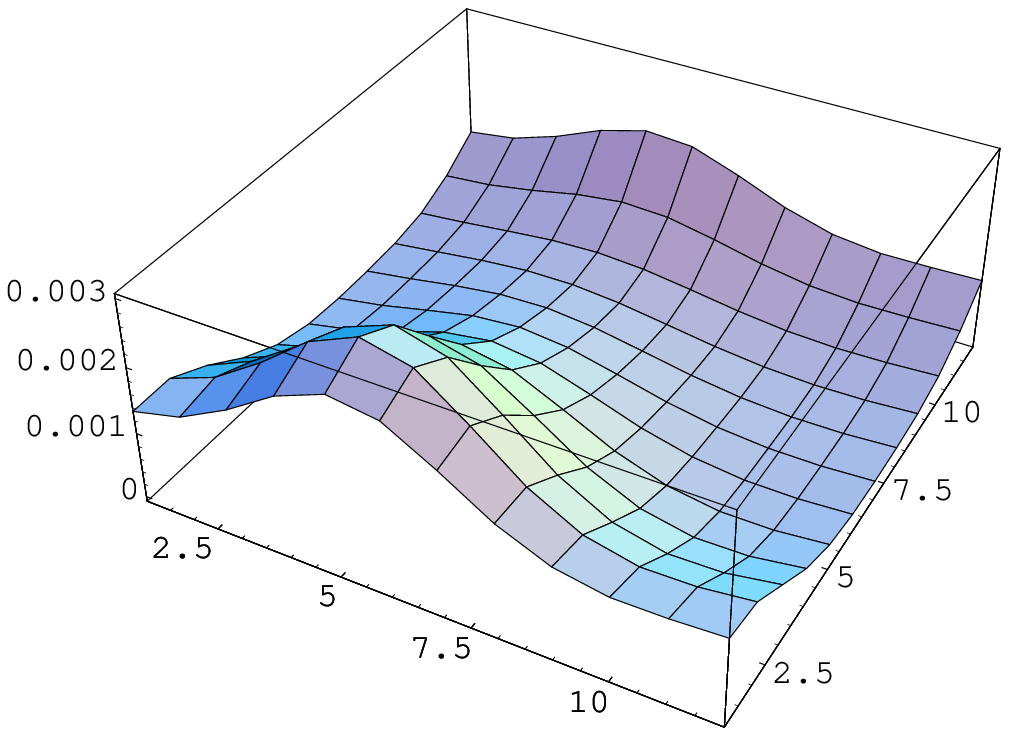, 
width=5cm, height=5cm}
\epsfig {file=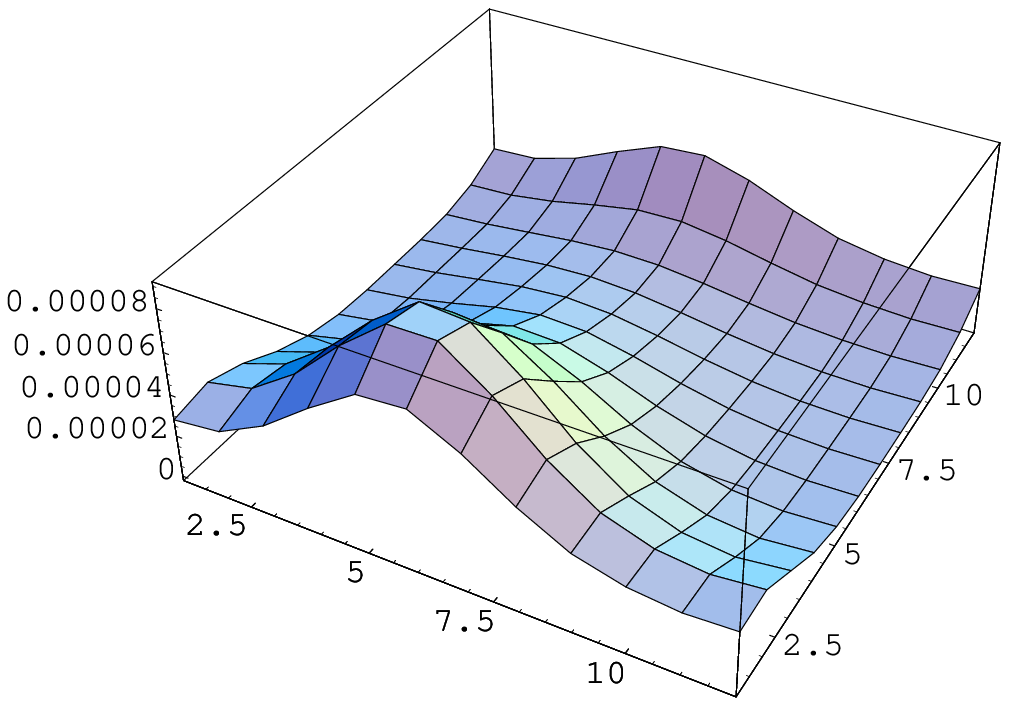, 
width=5cm, height=5cm} }
\caption{ \it Action (left) and chiral (right) densities for the $Q=2$ 
configuration; Both given  for  y-z plane.}
\label{fig_q=2}
\end{figure}
\end{enumerate}

If a (anti-)self-dual configuration consists of well separated instantons
or anti-instantons,  a new complication can arise. In the continuum, as  
the separation becomes infinite, the zero-modes tend to those of isolated 
instantons, one of which will satisfy the reality condition. This is one of 
the limiting cases that we mentioned before for which there is more than one
zero-mode satisfying the reality condition. It is clear that the chiral 
density of all states in this subspace of real zero-modes reproduces the 
sum of the action density of all the instantons weighted in various ways. As
the instantons approach, the reality only  remains true for a single (charge
conjugate pair) eigenmode  weighting all instantons equally: the
supersymmetric zero-mode.  
On the lattice,  corrections might spoil the picture for sufficiently 
separated instantons. The state which minimizes Eq.~(\ref{cond}) 
might have different weights for different well-separated structures. 
A signal that this is actually happening can be obtained from the
hierarchy of low-lying eigenvalues of $M$. A set of small eigenvalues 
separated by a large gap from higher ones is a clear indication of this 
situation. The supersymmetric solution  we are looking for is then a linear
combination of the states constructed from the eigenstates pertaining to 
the minimal eigenvalues of $M$. For sufficiently smooth configurations 
the correct  weights can be constrained since the contribution to the total 
action and topological charge of each separated structure must be in rational 
fraction of the total.
\newline 
We have looked at several  $Q=2$ configurations that 
contained well separated (anti) self-dual lumps. In one case we found
precisely the scenario  described above, while for other configurations the 
numerical method described in Section \ref{lattice} projected directly onto 
the  supersymmetric solution of Eq.~(\ref{zeromodes}).    
\newline
So far we have checked  that our numerical method of projecting onto the 
supersymmetric zero-mode Eq.~(\ref{zeromodes}) works well for (anti) self-dual
fields. In the next subsection we will consider a more complicated structure
 which no longer is a solution of the classical euclidean equations of motion.
 
\subsection{Other Smooth  Configurations}
\begin{figure}[ht]
\centerline {\epsfig {file=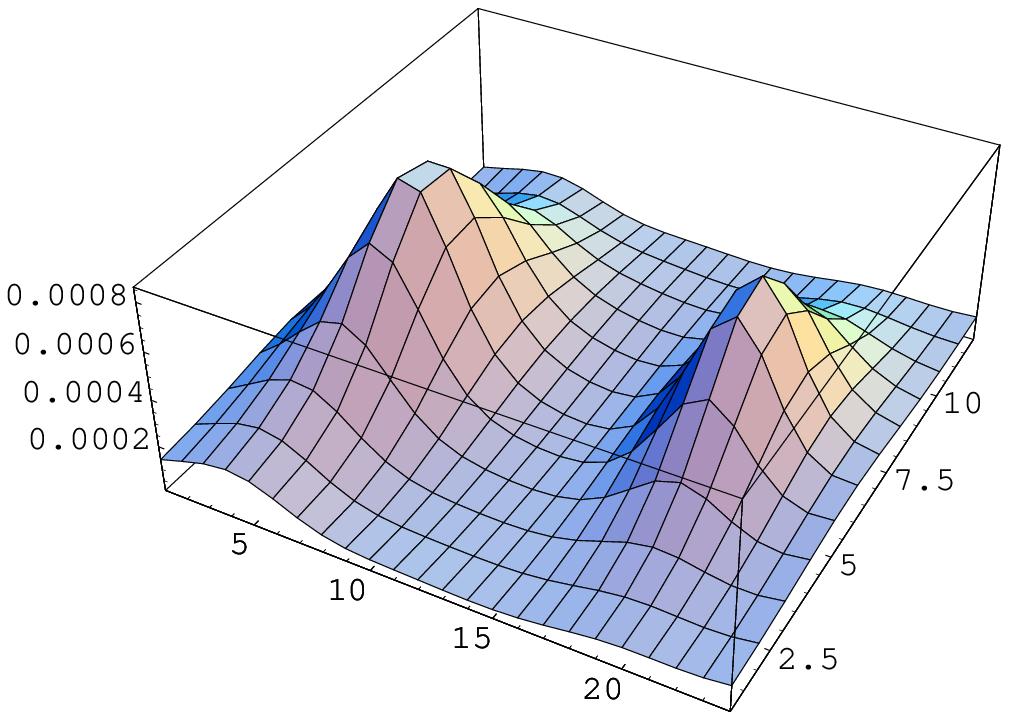 ,width=7cm, height=4.5cm}
\epsfig {file=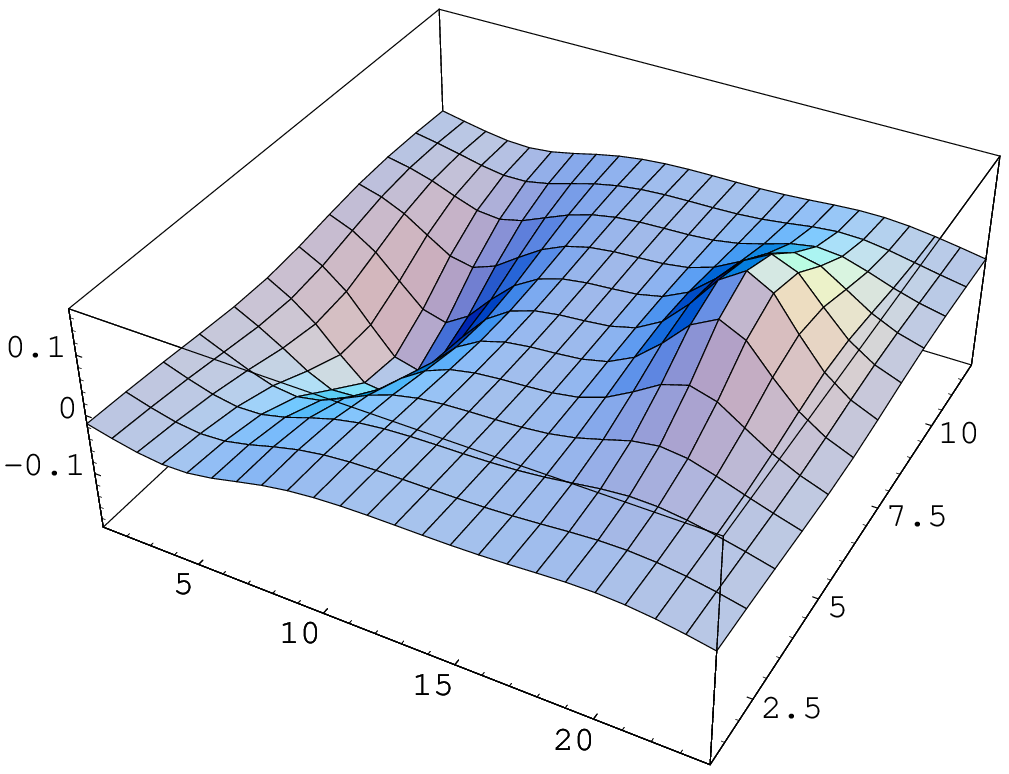, width=7cm, height=4.5cm}}
\caption{ \it Action density (left) and charge density (right) of the glued Q=0
configuration.}
\label{fig_q=0}
\end{figure}

\begin{figure}[ht]
\centerline {\epsfig {file=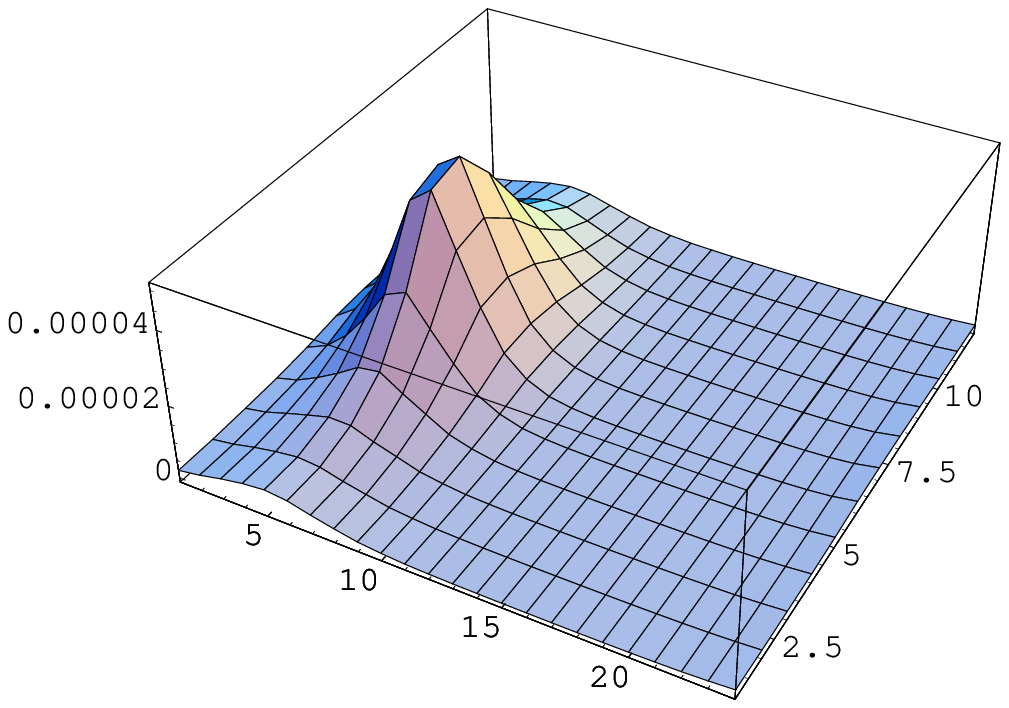 ,width=7cm, height=4.5cm}
\epsfig {file=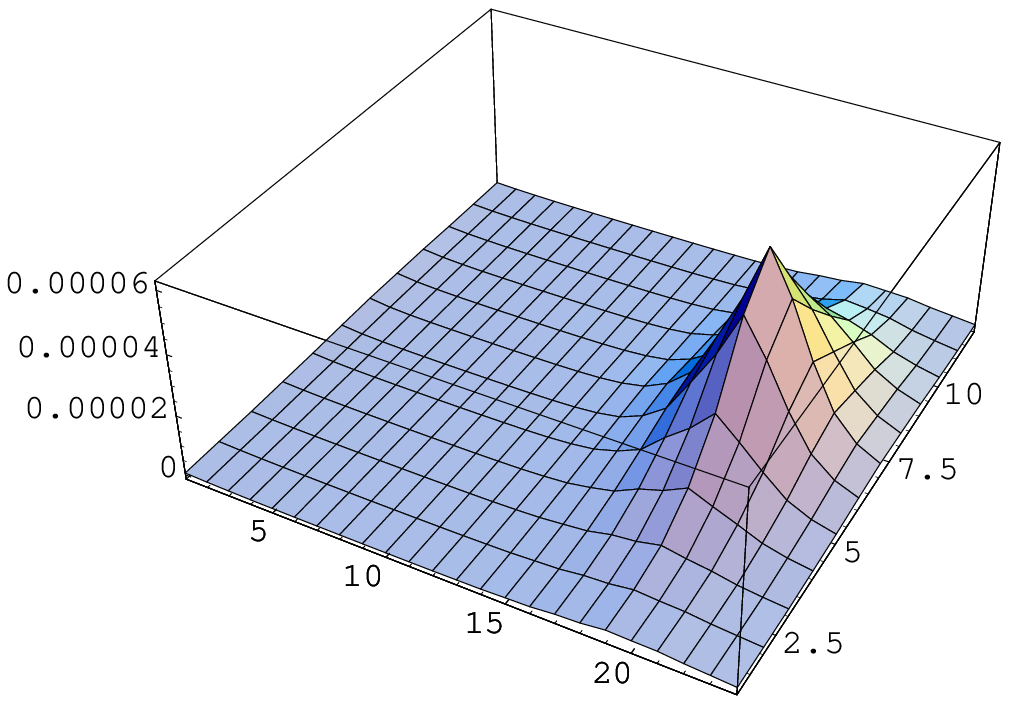, width=7cm, height=4.5cm}}
\caption{ \it  Chiral density $(\Psi^{-}_R)^{\dagger}\Psi^{-}_R$ (right) and
$(\Psi^{+}_L)^{\dagger}\Psi^{+}_L$ (left), for our instanton-anti-instanton
configuration.}
\label{fig_q=0_1}
\end{figure}

Our  next  step will be to  consider the case of smooth configurations which
are not solutions of the classical equations of motion. For that purpose 
we study a $Q=0$ configuration consisting of an instanton
and an anti-instanton, each carrying charge $|Q|=1$. Note that in this case 
we do not  have exact zero modes in the continuum. 
However, we expect that if the instantons are sufficiently 
separated, so that  their overlap is small,   the 
zero-modes pertaining to each instanton will give rise  to {\em near-zero} modes 
in the spectrum of $(\gamma_5D_W)^2$. 
In our  $Q=0$ instanton anti-instanton pair case, we should     find 
eight near-zero modes stemming from the four 
zero-modes of an isolated instanton. Furthermore, we 
expect to be able to identify  within  the space of {\em near-zero} modes 
one state in each chirality sector arising from the supersymmetric 
zero-modes Eq.~(\ref{zeromodes}) of the instanton and anti-instanton
separately. 
\newline
\newline
{\bf  $Q=+1-1=0$, m=k=\{000\} }
\newline
\newline
To produce  a smooth $Q=0$ lattice configuration consisting of an instanton 
anti-instanton
pair, we glued along the time direction the previously mentioned  $Q=-1$ 
configuration to a time reflected copy of itself. The resulting configuration 
has an increased  action density along the plane of gluing. We then applied 
several cooling steps to diminish the action. The resulting configuration 
 (see Figure \ref{fig_q=0}) has  an action density which looks indeed 
 like a superposition of an instanton and an anti-instanton. 
For this configuration we computed the twenty    
lowest-lying eigenmodes of the operator 
$(\gamma_5 D_W)^2$ at $r=0.3$ (see Table \ref{Q=0}).
\begin{table}[h]
\begin{center}
\begin{tabular}{| c | l | l | l |}
\hline
$\rm{Eigenvalue}$  &
\multicolumn{1}{|c|}{$\left < \phi, \gamma_5 \phi \right > $} &
$\rm{Eigenvalue}$  &
\multicolumn{1}{|c|}{$\left < \phi, \gamma_5 \phi \right > $} 
 \\
\hline
 $\lambda_1$=2.50 * $10^{-3}$ & -0.725 &  $\lambda_{11}$=1.32 * $10^{-2}$ &
 0.119  \\
\hline
 $\lambda_3$=3.42 * $10^{-3}$ & 0.337  &  $\lambda_{13}$=3.61 * $10^{-2}$ & 
-6.54 * $10^{-2}$  \\
\hline
 $\lambda_5$=4.94 * $10^{-3}$ & -0.439 &  $\lambda_{15}$=3.66 * $10^{-2}$  &
 6.70 * $10^{-2}$  \\
\hline
 $\lambda_7$=6.44 * $10^{-3}$ & 0.804  &  $\lambda_{17}$=5.20 * $10^{-2}$  &
 -4.67 * $10^{-2}$ \\
\hline
 $\lambda_9$=1.195 * $10^{-2}$ & -0.117 &  $\lambda_{19}$=5.26 * $10^{-2}$& 5.00 * $10^{-2}$ \\
\hline
\end{tabular}
\end{center}
\begin{center}
\parbox{12cm}{\caption{\label{Q=0} \it
Lowest lying eigenvalues of the operator $(\gamma_5 D_W)^2$ and chiralities 
of the respective eigenstates, for the instanton-antiinstanton configuration.}}
\end{center}
\end{table}
We see that in our case no clear gap in the eigenvalue spectrum  can be 
identified. The zero-modes of the original $Q=-1$ configuration 
(see Table \ref{Q=-1}) have been lifted to  near-zero modes.
Given the large separation of the instanton and anti-instanton, we expect that 
the near-zero modes associated to the supersymmetric states lie within the
eight lowest-lying modes. Hence, we computed the observables $S_\pm$
by minimizing expression (\ref{cond}) within this space. These quantities 
are shown in Figure (\ref{fig_q=0_1}). 
 We see that they reproduce nicely the shape of the self-dual and the anti self-dual parts  
of the action density respectively. We recalculated  $S_\pm$ within the full
space of  twenty  low-lying eigenmodes. Once more, our  results were found to be stable.

\subsection{Heated Configurations}
We now want to go one step further and investigate the efficiency 
of our method in the presence of quantum fluctuations. As explained 
in the introduction this amounts to the consideration of rough configurations.
Optimally we would like to add quantum fluctuations without distorting the 
long-wavelength topological structure of the configuration, so that we can 
compare our observable fields   $S_\pm$ with the  mentioned  structure.  
To achieve this goal  we use the following method. We begin with a classical
anti self-dual configuration for definiteness. Then  we apply a given 
number of heat-bath steps with a certain value of $\beta$. The local 
nature of the heat-bath algorithm\cite{HEATBATH} guarantees that the 
high-momentum modes will thermalize faster than the low momentum modes. 
Based on results obtained  in other contexts~\cite{ALLES93} we conclude
that a total of ten to twenty heating steps should be enough for this purpose.
The value of $\beta$  determines the typical size of the quantum 
fluctuations ( $\sim 1/\sqrt{\beta}$),  as well   as the effect of 
non-linearities. 
\newline
Thus, we started  by considering a very high value of $\beta$, 
namely $\beta=22$. This introduces  gaussian perturbations (up to 
gauge equivalence). As initial configuration, we chose the second one in  
Table \ref{conf_sum}. We  applied a total of twenty heating steps to it. 
After each heating step,  we calculated the sixteen lowest-lying 
eigenstates of $(\gamma_5 D_W)^2$ at $r=0.3$. 
From them, we  calculated the observable $S_-$ according
to the procedure outlined in Section \ref{projection}.
The results obtained confirm the good behavior of  $S_-$ with 
respect to quantum fluctuations. At each heating step $\beta$,
$S_-$ resembled closely the shape of the starting configuration. This 
contrasts with the evaluation of the action density itself (or its 
anti-self-dual part) which is dominated by noise, masking completely  
the initial topological structure. Indeed, the action density after 
heating is two orders of magnitude higher than the action density of the
original field.
\newline
Then  we considered  two smaller values of $\beta$, $\beta=5$ and $\beta=2.57$.
The latter lies in the relevant scaling region of Monte-Carlo simulations.
As we lower $\beta$ the fluctuations have a larger size and we depart from
the gaussian situation, in which analytical control is still possible. 
Furthermore, the probability of generating typical lattice artifact effects,
like the dislocations \cite{LUSCHER82,PUGHTEPER_1}, increases. These
structures can contribute sizably to the topological charge for some lattice
definitions of the quantity. This effect on the topology can also show 
up in the Dirac spectrum.   It has been shown in $SU(2)$ gauge 
theory that, with the plaquette action, dislocations appear quite frequently,
but that they disappear after only a few cooling steps \cite{PUGHTEPER_1,PUGHTEPER_2}.  
For this reason we introduce a small number of cooling steps (3) on the heated configuration 
and proceed with our method on this slightly cooled configuration.
The use of a fixed and small number of cooling steps might be acceptable 
if we can show that the results are insensitive to the details of their
implementation.
\newline
For that purpose  we have also used   APE-smearing \cite{APE} 
(see Section \ref{technical}) instead of cooling to smoothen the heated 
configurations. We have chosen $N=8$ and $c=0.45$ as smearing parameters, 
so that the effective smearing radius and 
the number of cooling steps are of the same magnitude.
As we will show  below both methods do not only give consistent 
results, but lead to values of $S_\pm$ that are strikingly similar.

\subsubsection{Updating at $\beta=5.00$}
\begin{figure}
\centerline {\epsfig {file=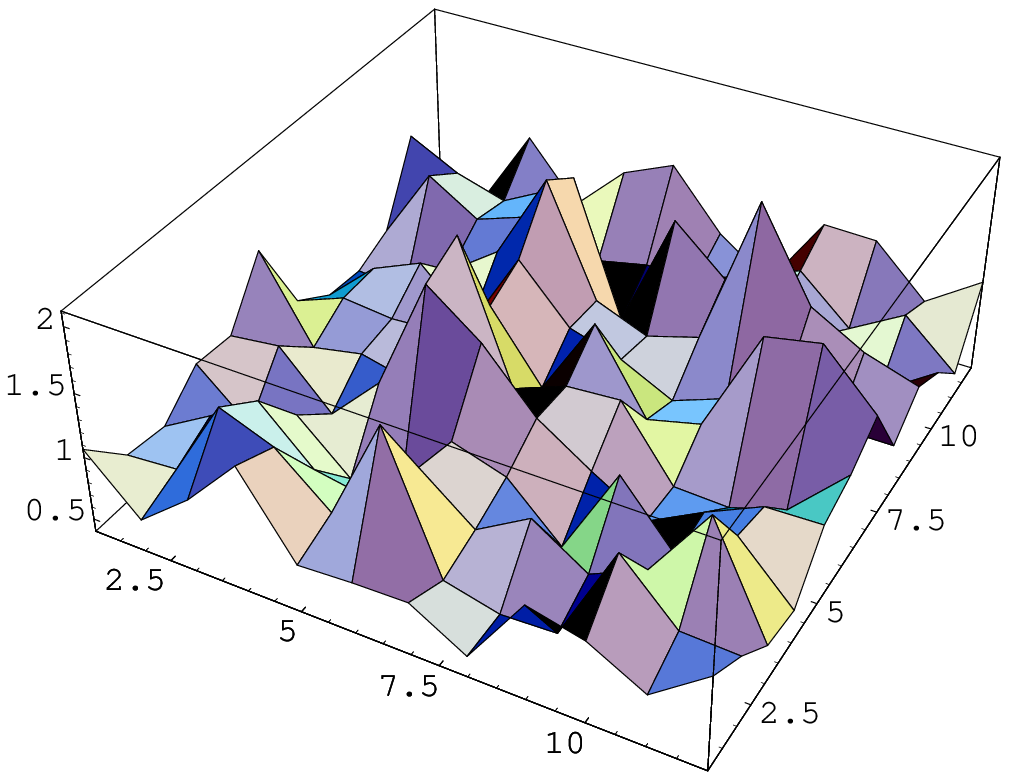 ,width=5cm, height=5cm}
\epsfig {file=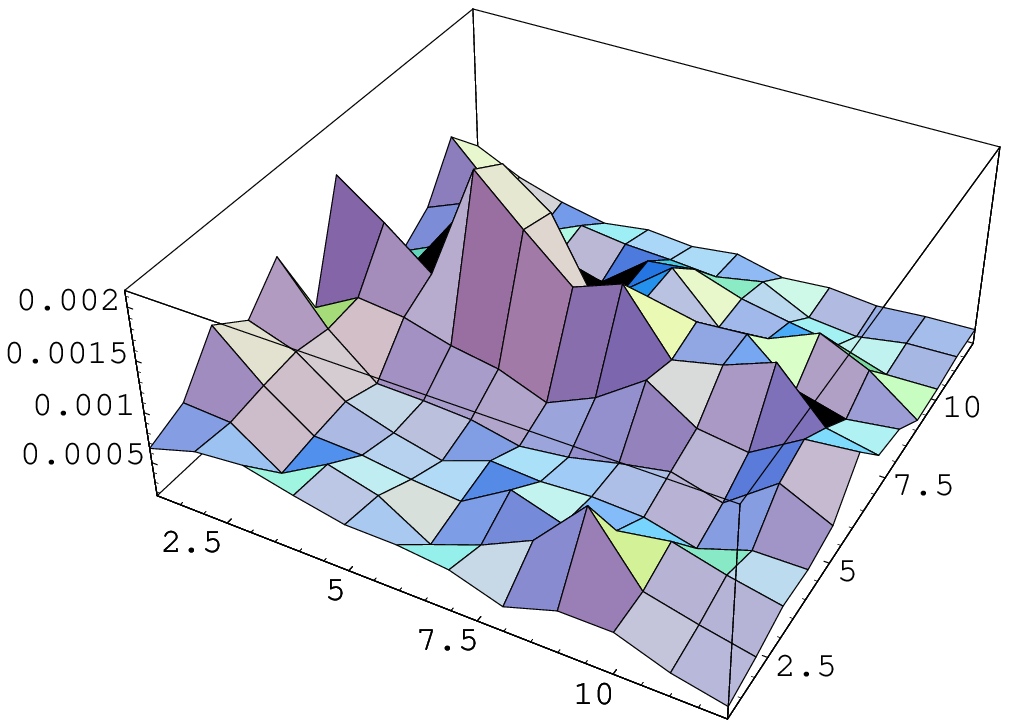, width=5cm, height=5cm}
\epsfig {file=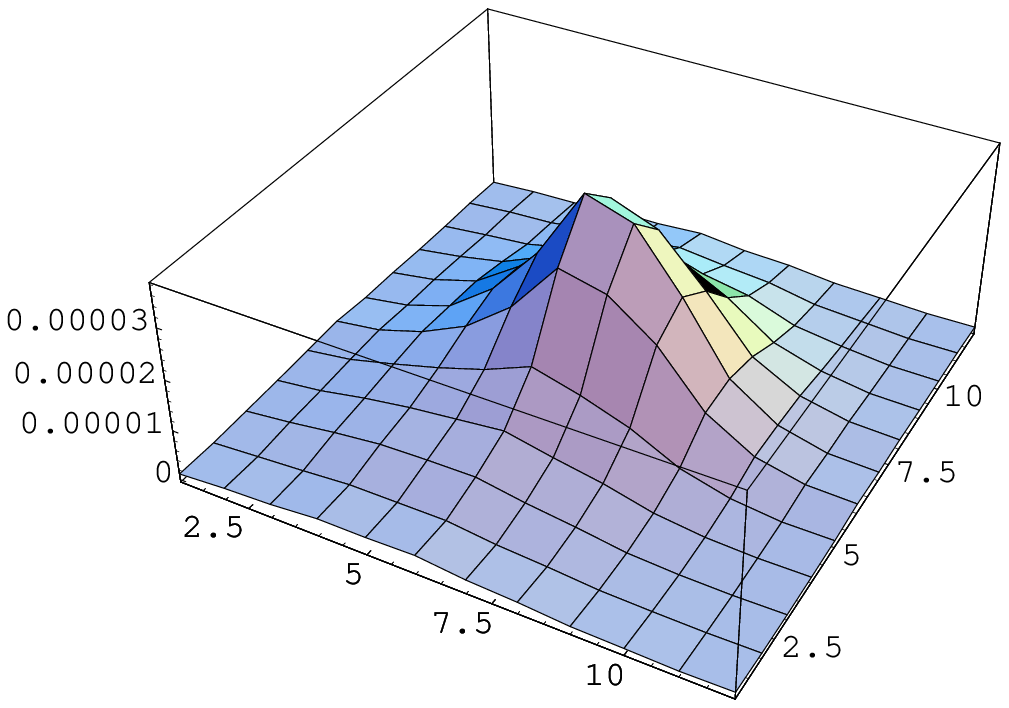,width=5cm, height=5cm}}
\caption{ \label{fig_q=-1_beta5_step5} \it Action density of the heated 
configuration after 5 heating 
steps (left) with $\beta=5.00$, and after applying 3 cooling steps to the
previous one(middle). The right figure is the chiral density 
$\Psi_R^{\dagger} \Psi_R$  obtained from the near  zero-modes of the cooled
configuration.}
\end{figure}

\begin{figure}
\vspace{1cm}
\centerline {\epsfig {file=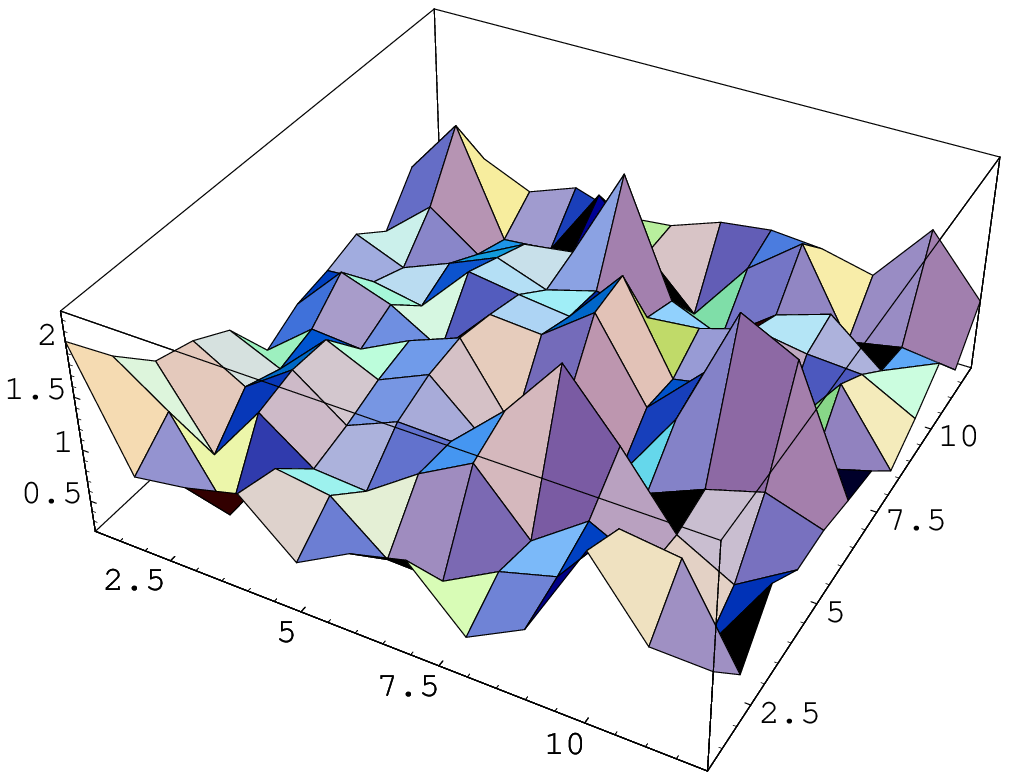 ,width=5cm, height=5cm}
\epsfig {file=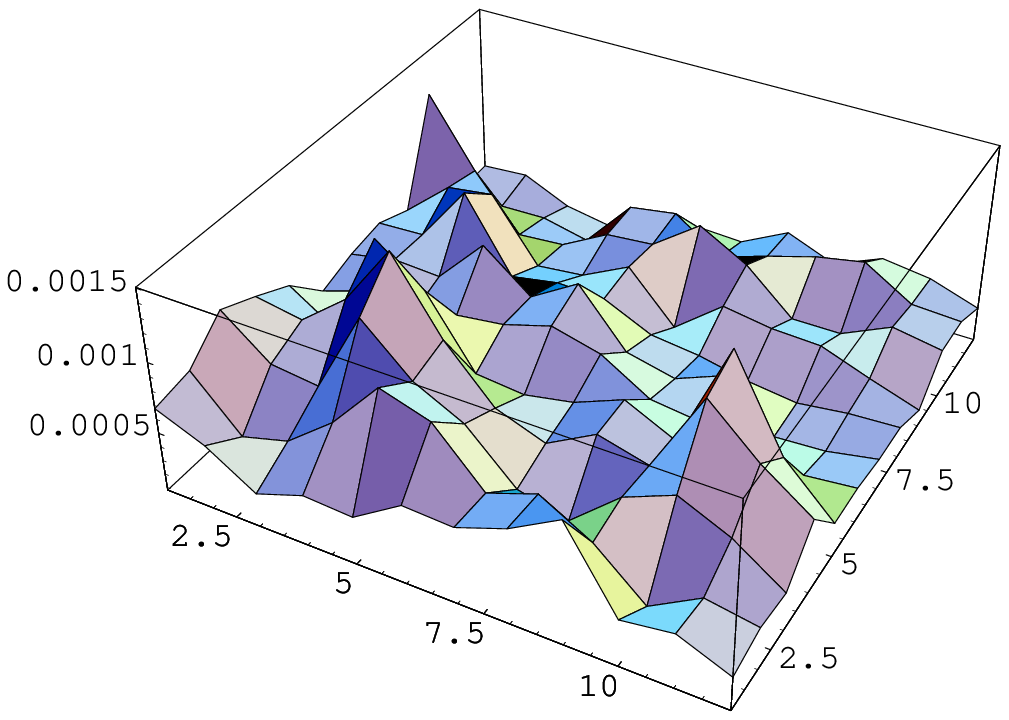, width=5cm, height=5cm}
\epsfig {file=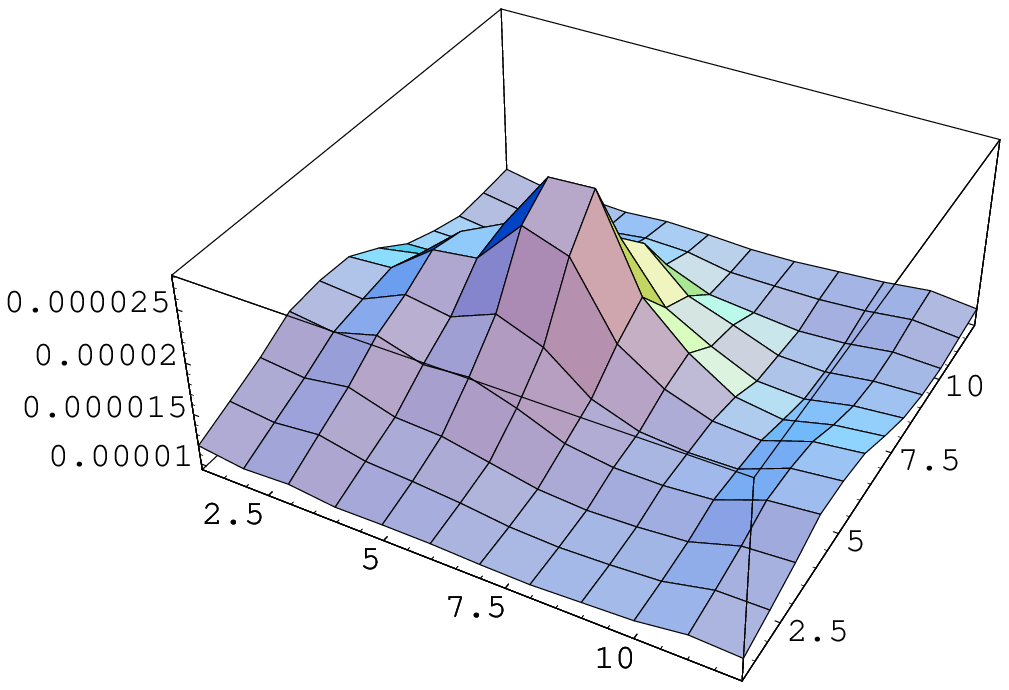,width=5cm, height=5cm}}
\caption{ \label{fig_q=-1_beta5_step20} \it 
The same quantities as in the previous figure 
but  after 20 heating steps at $\beta=5.00$.}
\end{figure}

Again we started with the $Q=-1$ configuration appearing as 
example two in Subsection \ref{sd_conf}, with action density 
displayed in Figure \ref{fig_q=-1}.
At $\beta=5.00$ we subjected the configuration to twenty heating steps. Every
fifth step we produced a (three times) cooled or a ($N=8$ $c=0.45$) smeared 
configuration.  The corresponding values of
of the topological charge and the action  can be found
in Table \ref{heatcool_beta5}.
As we can see   no extra charge seems to have been created during heating.
\begin{table}[h]
\begin{center}
\begin{tabular}{| c | l | l | l | l | l | l |}
\hline
$\rm{heating}$&
\multicolumn{1}{|c|}{$S/(8\pi^2)$} &
\multicolumn{1}{|c|}{$Q_L$} &
\multicolumn{1}{|c|}{$S_{cool}$} &
\multicolumn{1}{|c|}{$Q_{L,cool}$} &
\multicolumn{1}{|c|}{$S_{smear}$} &
\multicolumn{1}{|c|}{$Q_{L,smear}$} \\
\hline
 0 & 1.02  & -0.90  & --  & --  & --  & --  \\
\hline
 5 & 991.8 & -0.56 & 6.5 & -0.87 & 5.0 & -0.87  \\
\hline
 10 & 993.9 & -0.41 & 6.6 & -0.86 & 5.2 & -0.84 \\
\hline
 15 & 990.3 & -1.02 & 6.6 & -0.86 & 5.1 & -0.86 \\
\hline
 20 & 993.9 & -0.60 & 6.5 & -0.86 & 5.0 & -0.86 \\
\hline
\end{tabular}
\end{center}
\begin{center}
\parbox{12cm}{\caption{\label{heatcool_beta5} \it
Values of the action $S$ (divided by $8\pi^2$) and topological charge (field theoretic) of the 
heated (at $\beta=5.00$), cooled (3 times) or smeared configurations on a $12^4$ lattice.}}
\end{center}
\end{table}
\newline
For each of the cited configurations the sixteen
lowest lying eigenmodes of the operator $(\gamma_5D_W)^2$ have been 
calculated at $r=0.3$. We then determined $S_\pm$ by applying the minimization 
of expression (\ref{cond}) in this sixteen dimensional subspace of eigenmodes.
The results  for  5 and 20 heating steps can be found in Figure 
(\ref{fig_q=-1_beta5_step5}, \ref{fig_q=-1_beta5_step20}).

We observe that in all cases the shape of $S_-$  resembles  fairly 
well the action density of the original configuration. After twenty heating
steps the qualitative agreement is still quite good. 
Similar results are obtained for  the smeared configurations.
Hence, we conclude that our results are independent of the method 
used --- smearing or cooling --- to remove the lattice artefacts.
\newline

It is interesting to point out that the heating process does not seem to 
have changed the long wavelength structure of the initial configuration. 
This is due to the large value of $\beta$ and the small number of 
heating steps. In the next subsection we will see that the situation changes 
for smaller values of $\beta$. 
\subsubsection{Updating at $\beta=2.57$}
\begin{figure}
\centerline {\epsfig {file=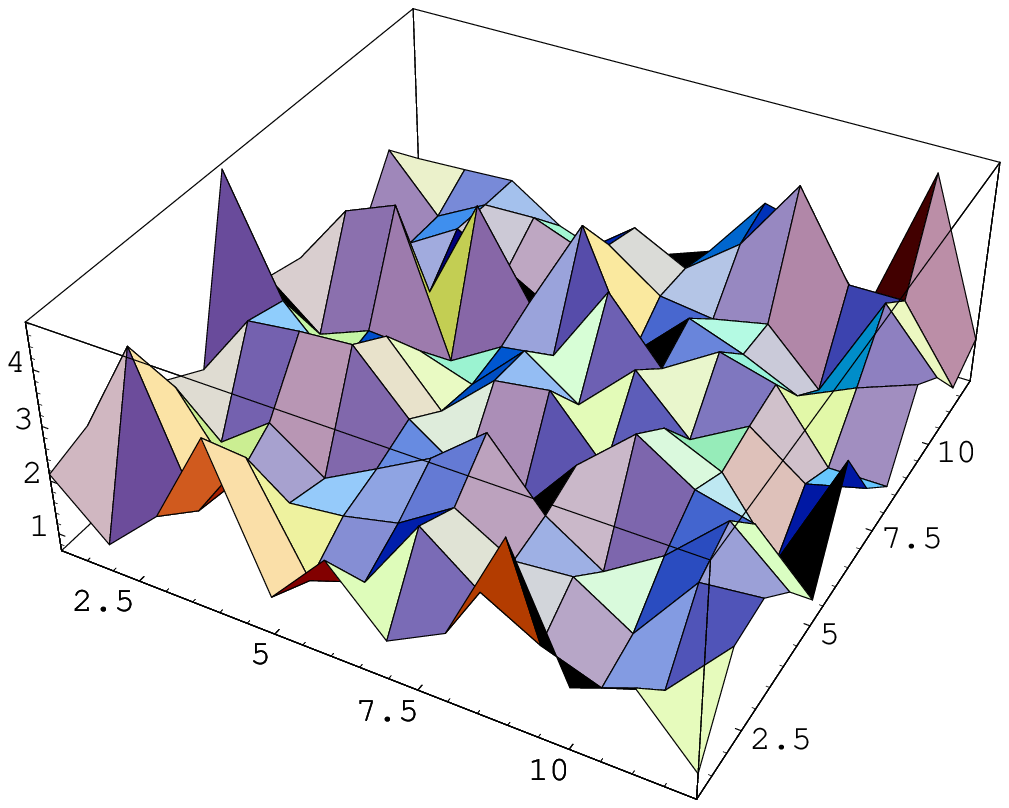 ,width=5cm, height=5cm}
\epsfig {file=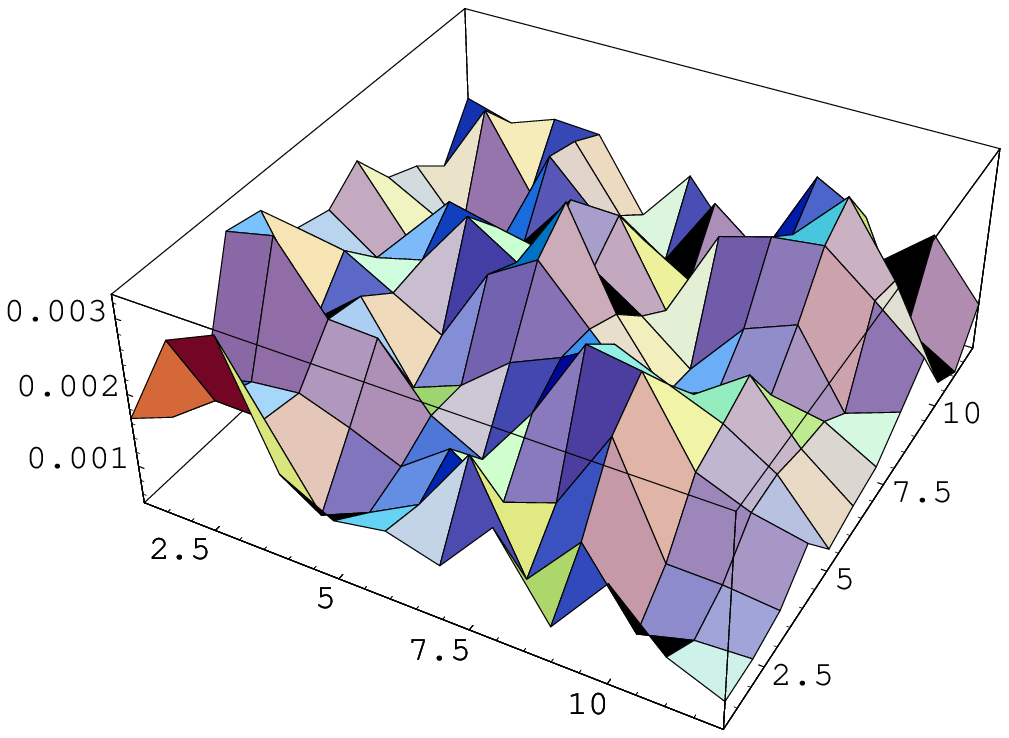, width=5cm, height=5cm}
\epsfig {file=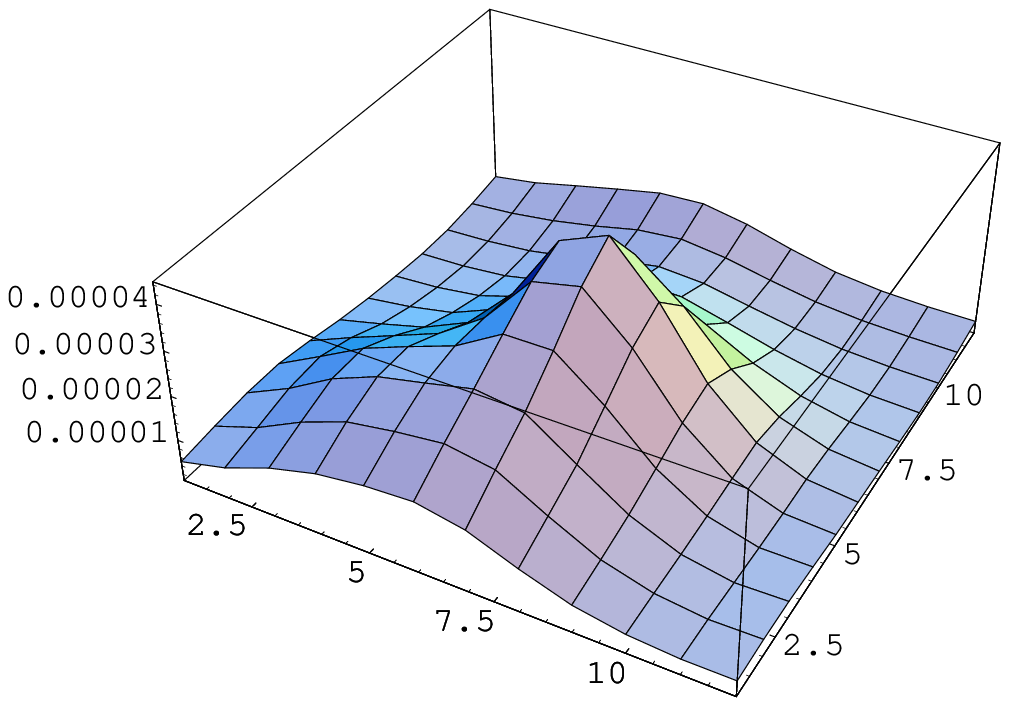,width=5cm, height=5cm}}
\caption{ \label{fig_q=-1_beta2.57_step5} \it 
The same as Fig.~\ref{fig_q=-1_beta5_step5} but after 5 heating steps at
$\beta=2.57$}
\end{figure}

\begin{figure}
\centerline {\epsfig {file=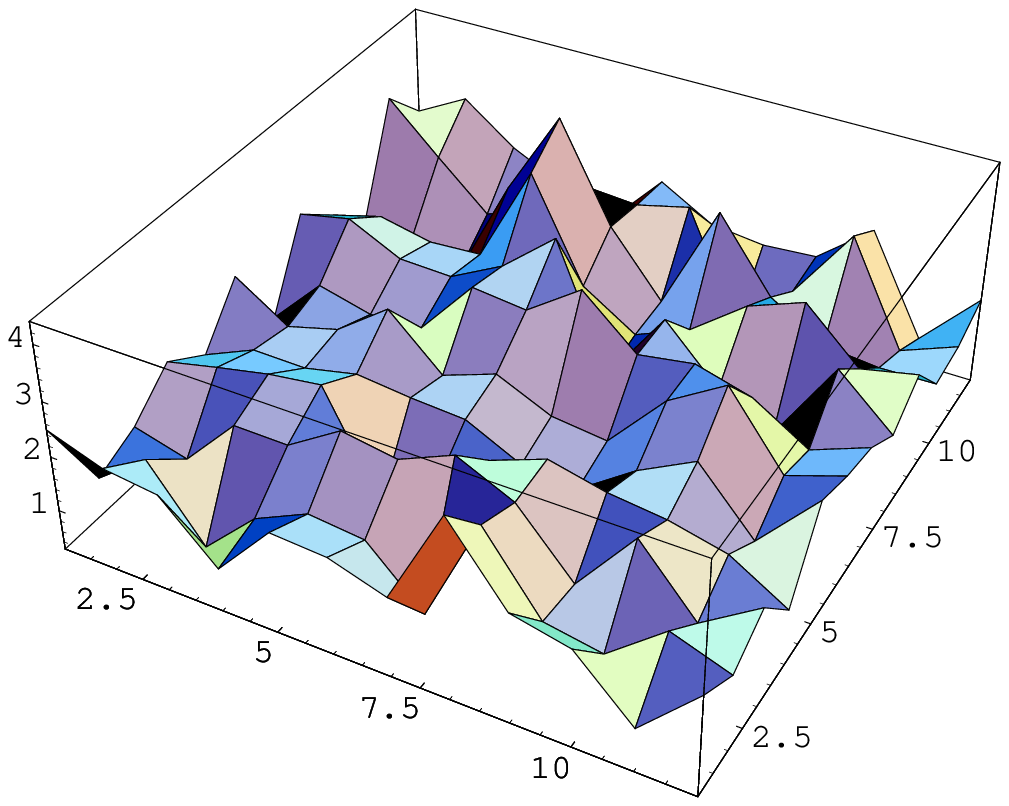 ,width=5cm, height=5cm}
\epsfig {file=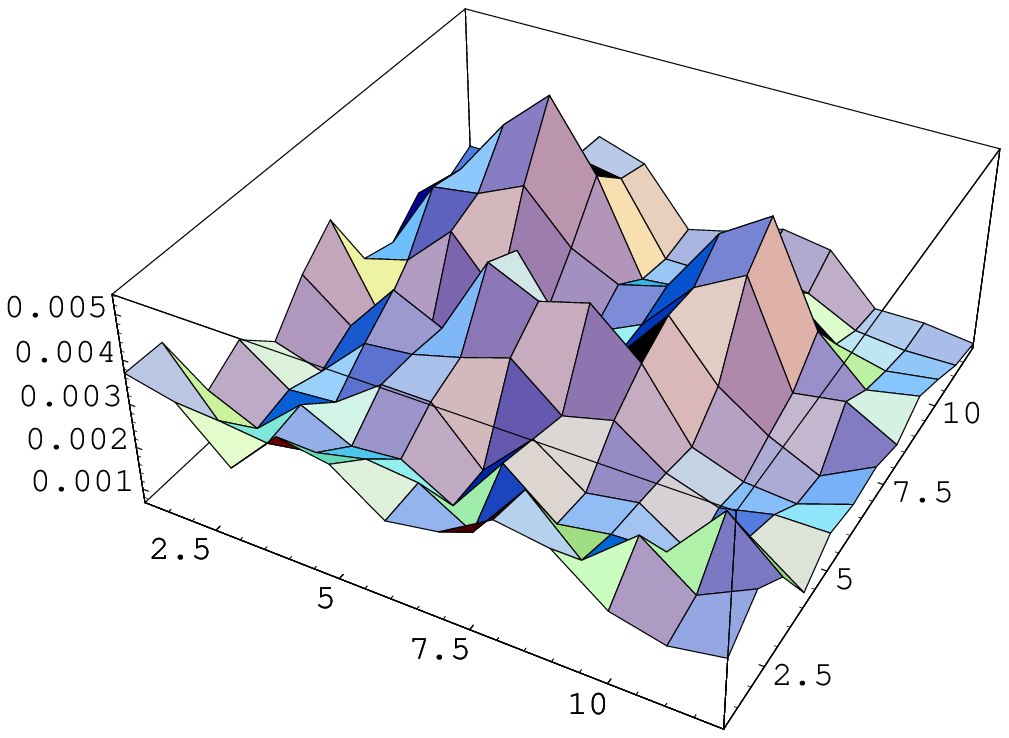, width=5cm, height=5cm}
\epsfig {file=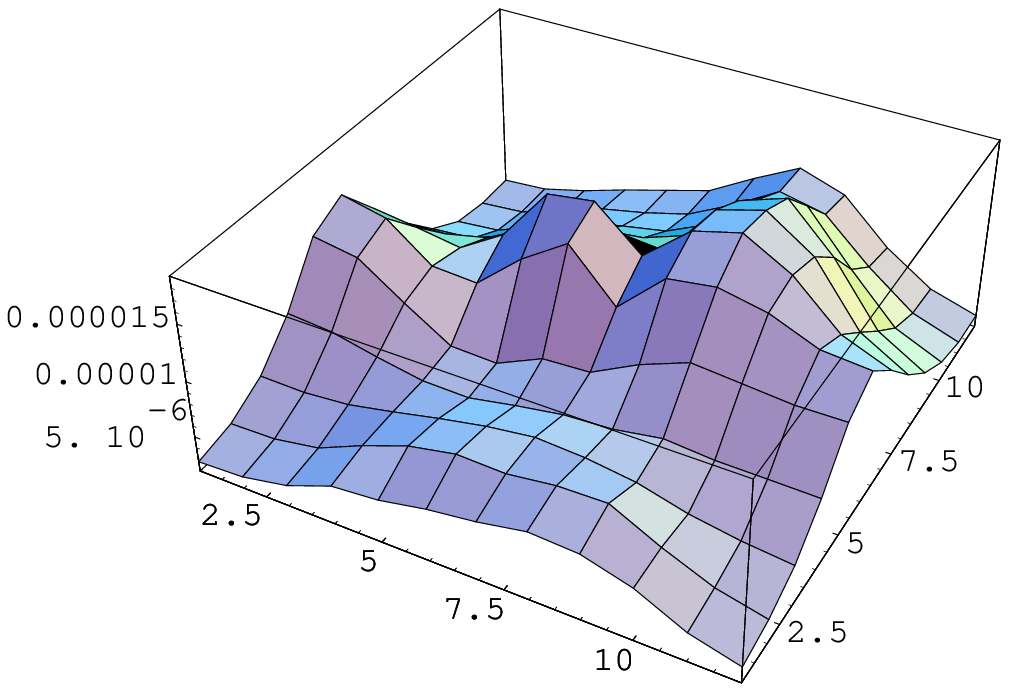,width=5cm, height=5cm}}
\caption{ \label{fig_q=-1_beta2.57_step20} \it The same as
Fig.~\ref{fig_q=-1_beta5_step20} for $\beta=2.57$}
\end{figure}

Let us now turn to the  $\beta=2.57$ case,  which is the most interesting 
case in understanding the behavior of our observables previous to their use
for Monte Carlo generated configurations. Our procedure is identical to 
the previous case. Results for the  topological charge and the action 
 can be found in Table \ref{heatcool_beta2.57}.
 \newline 
 At this value of $\beta$ we expect that, after a sufficient number of 
 heating steps are applied, the configuration will have modified the 
 long-wavelength structure of the initial configuration. It is then 
 unclear what the shape of $S_\pm$ has to be. 
\begin{table}[h]
\begin{center}
\begin{tabular}{| c | l | l | l | l | l | l |}
\hline
$\rm{heating}$&
\multicolumn{1}{|c|}{$S/(8\pi^2)$} &
\multicolumn{1}{|c|}{$Q_L$} &
\multicolumn{1}{|c|}{$S_{cool}$} &
\multicolumn{1}{|c|}{$Q_{L,cool}$} &
\multicolumn{1}{|c|}{$S_{smear}$} &
\multicolumn{1}{|c|}{$Q_{L,smear}$} \\
\hline
 0 & 1.02  & -0.90 & --  & --  & --  & --\\
\hline
 5 & 2074.7 & -0.96 & 16.6 & -0.88 & 14.56 & -0.88 \\
\hline
 10 & 2119.7 & 0.47 & 21.8 & -0.84 & 21.48 & -0.84 \\
\hline
 15 & 2110.7 & 1.38 & 23.5 & -0.89 & 23.46 & -0.86 \\
\hline
 20 & 2110.2 &-0.84 & 22.7 & -0.91 & 22.85 & -1.00 \\
\hline
\end{tabular}
\end{center}
\begin{center}
\parbox{12cm}{\caption{\label{heatcool_beta2.57} \it
Action and topological charge (field theoretic) of the heated (at
$\beta=2.57$), cooled and 
smeared configurations on a $12^4$ lattice.}}
\end{center}
\end{table}
Notice, however,  that the value of $Q_L$ after cooling indicates that no 
extra overall charge has been created during the updating. 
This does not exclude  that a number of instanton-anti-instanton pairs 
have been created in this process.
\newline
As in the previous case, we compute the fields $S_\pm$ in terms of 
the   low-lying eigenmodes of the operator $(\gamma_5D_W)^2$ after 5 
and  20 heating steps. Results are shown in  Figs. 
(\ref{fig_q=-1_beta2.57_step5} - \ref{fig_q=-1_beta2.57_step20}). 

\begin{figure}
\centerline {\epsfig {file=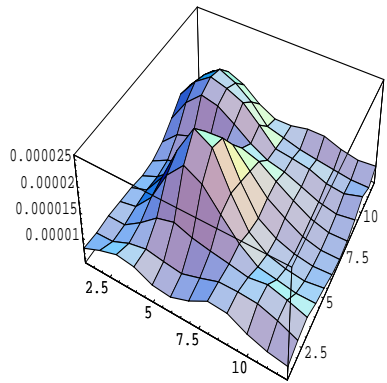,width=5cm, height=5cm}}
\caption{ \label{fig_q=-1_beta2.57_step20_right} \it Chiral density 
$\Psi_L^{\dagger} \Psi_L$ of minimized linear combination obtained from
the cooled
configuration at $\beta=2.57$.}
\end{figure}

We observe, that after 5 heating steps one can still recognize in $S_-$ the
structure of the original anti-instanton.  After 20 heating steps,
however, we can see a richer structure emerging. The main peak of the 
original instanton is still present, but two additional lumps  in
this plane  seem to 
have been created during the heating process. Since there seems to be 
no net creation of topological  charge, we expect to find  two structures 
in $S_+$ corresponding to positively charged lumps.
The result is shown  Fig.~\ref{fig_q=-1_beta2.57_step20_right}, and we can clearly identify two isolated structures, which we 
associate to two instantons that have been created during the heating process.

\begin{figure}
\centerline {\epsfig {file=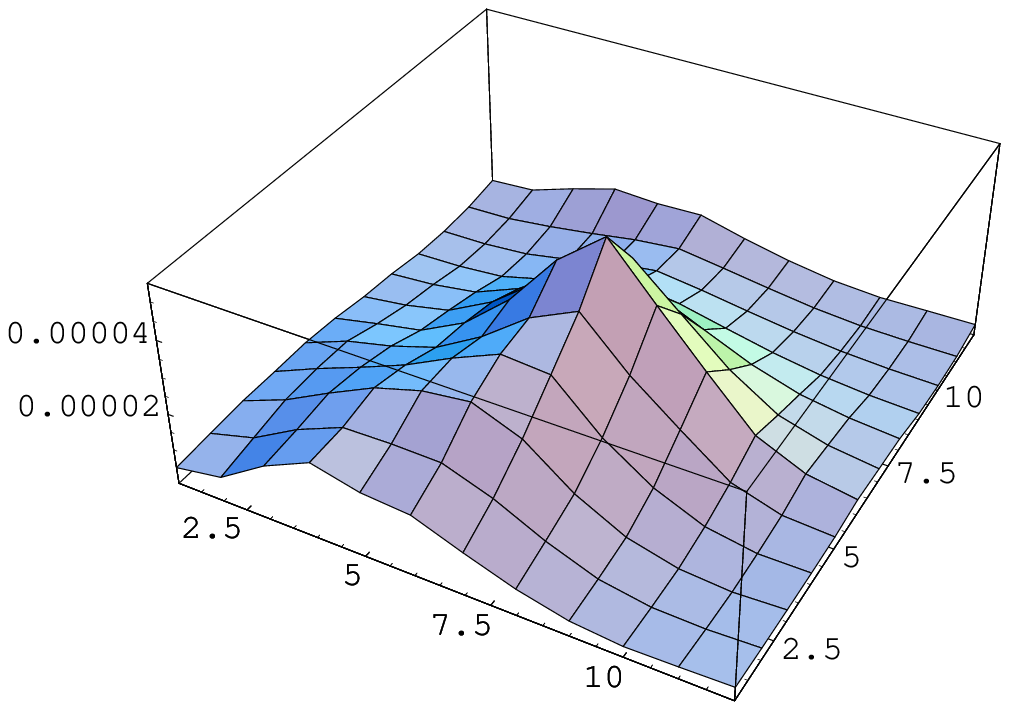, width=5cm, height=5cm}
\epsfig {file=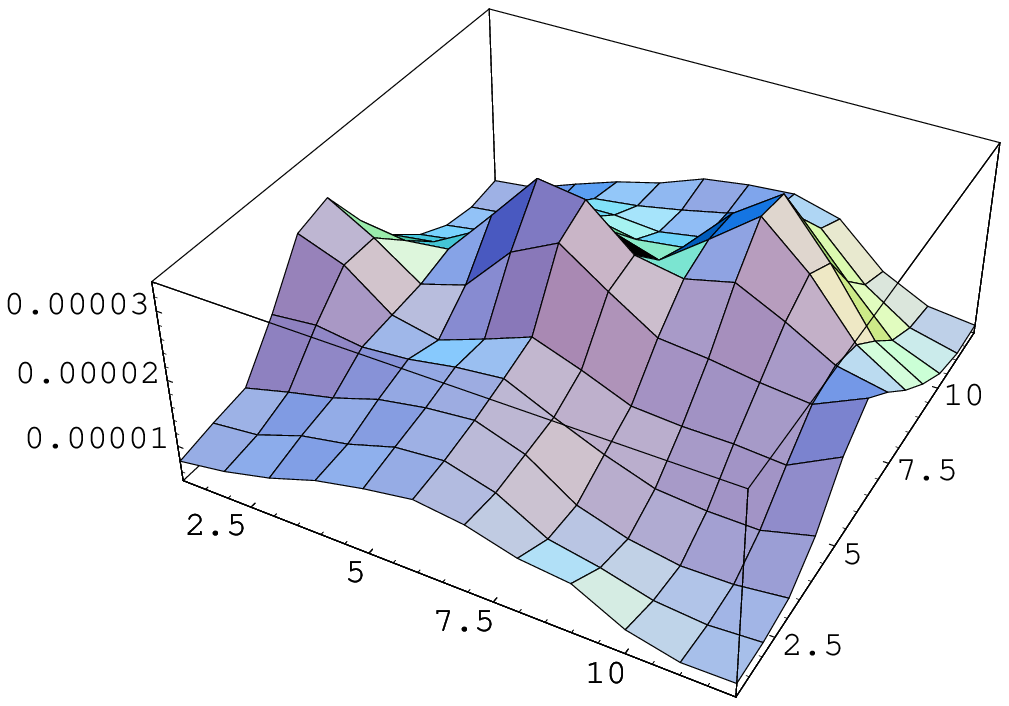,width=5cm, height=5cm}}
\caption{ \label{fig_q=-1_beta2.57_step20_smear} \it Chiral density 
$\Psi_R^{\dagger} \Psi_R$ (left) of minimized linear combination obtained from
the smeared
configurations after 5 heating steps (left) and after 
20 heating steps (right).}
\end{figure}

Let us compare the chiral densities obtained from the cooled configurations 
to the ones that result by carrying out the same procedure on the smeared
configurations.
The corresponding chiral densities for the heated and successively smeared 
configurations can be found in Figure \ref{fig_q=-1_beta2.57_step20_smear}.  
These figures when compared to the chiral densities obtained from the cooled 
configurations are strikingly similar. Accordingly we conclude that 
smearing as well as cooling disposes of the lattice artefacts in the same 
manner and that the same underlying structure is revealed by our procedure.

\section{Conclusions}

In this work we have proposed a new tool to investigate the vacuum  
structure of  gauge field theory. It is in the spirit of using fermionic
modes as probes of gauge field structure. However, we make use of 
fermions in the adjoint representation. For gauge fields which are 
exact solutions of the euclidean equations of motion, Supersymmetry 
gives rise  to a direct connection between the structure of some zero-modes and the 
gauge field strength. In the presence of quantum fluctuations one can use 
this idea to construct observables $S_\pm$ which reflect the underlying 
self-dual and anti-self dual components and are better behaved in the 
ultraviolet. 

In this paper our main goal has been to present the ideas 
and to test its ability to reconstruct the underlying gauge field structure
in an increasingly complex but  controlled situation. 
Thus, we have first analyzed the case of smooth self-dual gauge fields 
and shown that indeed our method  is capable of extracting the
supersymmetric zero-modes that match the shape of the action density. 
Next we have studied the case of smooth, non-self-dual configurations. 
Finally we have analyzed configurations having controlled quantum fluctuations.
This was achieved by applying a number of heat-bath sweeps to an initially 
self-dual configuration. We used the number of heating steps and the value of 
 $\beta$ to monitor the size of quantum fluctuations and their local nature.
For small enough number of heating steps or large enough $\beta$,
the chiral density of the linear combination of low-lying eigenmodes that best
approximates the supersymmetric mode, reproduces the action density of the
underlying anti-instanton on which heating was performed.
This contrasts with the value of the action 
density itself, which has little resemblance to the initial structure.

For $\beta=2.57$ we have found that after 20 heating steps, next to the
original anti-instanton, additional structures have developed during the heating
process. We interpret them as instanton anti-instanton pairs that have been created
during updating. Although  we had to apply three cooling steps to the
configuration before analyzing the zero-modes, we argue that this operation
has not distorted the underlying structure that we are looking for. In support of 
this claim we showed  that the shape  obtained is remarkably similar to
the one gotten from the Ape-smeared configuration instead.

The analysis presented in this paper serves as a test of  the 
usefulness of our construction in being able to unravel, for a given gauge 
field  configuration, the underlying topological structure masked by short 
wavelength fluctuations. The next step should be that of addressing Monte
Carlo generated configurations directly. For that purpose it is presumably
essential to use a lattice Dirac operator, such as Neuberger operator, 
which preserves an exact notion of chirality. 

We hope our work provides a  new tool that can be used, by itself or in 
conjunction with other methods, to study the topological structures present
in the QCD vacuum in the spirit of the works mentioned in the 
introduction~\cite{DeGrand01,DeGrand02,Horvath02a,Gattringer01,Hip02,Edwards02,
Blum02,Gattringer01_2,Horvath02b,Horvath03,Horvath04,Horvath05,Gattringer03,
Gattringer04}.

\section*{Acknowledgements}We thank Margarita Garc\'{\i}a  P\'erez for  
interesting discussions, suggestions  and a critical reading of the manuscript.

\newpage
\section{Appendix}

%
%
\label{APPCONVENTIONS}
In this Appendix we list the conventions and definitions that were used 
throughout this work.
\newline
\\
\large \textbf{Generalities} \normalsize
\newline
\\
The number of colors has been denoted by $N$. For the specific 
numerical 
calculations we work in the adjoint representation of SU(2), hence $N=2$.
Color in the adjoint representation is labeled by Latin lower indices of the 
form $a,b,c...=1,...,N^2-1$.
\newline
Greek lower case letters from the beginning of 
the alphabet such as $ \alpha,\beta,...=1,...,4$ label the spinorial indices.
Greek lower case letters from the middle of the alphabet such as 
$\mu,\nu,\rho,\sigma,...=0,...,3$ denote Lorentz space-time components 
(in Euclidean space-time).
\newline
\\
\large \textbf{SU(N) group} \normalsize
\newline
\\
The generators  of the Lie-algebra $su(N)$ of the gauge group $SU(N)$ 
in the fundamental representation are
denoted by $T^a$ with $a \in \{1,...,N^2-1 \}$ and are given by   $N
\times N$ matrices.
Their commutation relations  read
\begin{equation}
[T^a,T^b]=if^{abc}T^c.
\end{equation}
where $f^{abc}$ are the structure constants of the group. 
These generators are normalized according to 
\begin{equation}
\nonumber
{\rm Tr} [T^a T^b]=\frac{1}{2}\delta^{ab}.
\end{equation}
For a given group element $g \in SU(N)$, $U(g)$ will denote its corresponding 
matrix in the fundamental representation and $V(g)$ in the adjoint
representation. The matrix elements of the latter can be expressed in terms
of $U(g)$  as follows
\begin{equation}
V(g)^{ab}=\frac{1}{2}{\rm Tr}[U^{\dagger}(g)T^aU(g)T^b].
\label{adj_link}
\end{equation}
\large \textbf{Pauli and Euclidean Dirac matrices} \normalsize
\newline
\\
The symbol $\tau^a$ denotes the Pauli matrices:
\begin{equation}
\tau_1=\left(
\begin{array}{cc}
0 & 1 \\
1 & 0 
\end{array} \right),
\ \ \tau_2=\left(
\begin{array}{cr}
0 & -i \\
i & 0 
\end{array} \right),
\ \ \tau_3=\left(
\begin{array}{cr}
1 & 0 \\
0 & -1 
\end{array} \right). 
\end{equation}
We also define the quaternionic basis $\sigma_{\mu}= (1\!\!1, -i\vec{\tau})$, 
$\overline{\sigma}_{\mu}=\sigma_{\mu}^{\dagger}=(1\!\!1, i\vec{\tau})$.
\newline
The hermitian Dirac matrices $\gamma_{\mu}$ in the Weyl representation in
Euclidean space-time are given in terms of the previous matrices by
\begin{equation}
\gamma_{\mu}= \left(
\begin{array}{rc}
0 & \sigma_{\mu} \\
\overline{\sigma}_{\mu} & 0
\end{array}
 \right).
\end{equation}
We also define the following matrices
\begin{equation}
\gamma_5=\gamma_1\gamma_2\gamma_3\gamma_0=\left(
\begin{array}{cc}
1\!\!1 & 0 \\
0 & -1\!\!1 
\end{array}
\right),
\ \ \ \
C=\gamma_0\gamma_2=\left(
\begin{array}{cc}
i\tau_2 & 0 \\
0 & -i \tau_2
\end{array} \right),
\end{equation}
The charge conjugation matrix $C$ fulfills the following relations
\begin{equation}
C^{-1}=-C=C^{T}, \ \ \gamma_{\mu}^T=-C^{-1}\gamma_{\mu}C, \ \
\gamma_5=C^{-1}\gamma_5C.
\end{equation}
The eigenstates of $\gamma_5$ with eigenvalue $1$ are called positive
chirality of left-handed spinors. Conversely, negative chirality or
right-handed spinors apply for the eigenstates of eigenvalue $-1$. 
Any spinor can be decomposed into a sum of a right-handed and a left-handed
spinor  $\psi=\psi_R+\psi_L$,  where the latter can be obtained from $\psi$
using the projection operators $P_\pm= (1\!\!1 \pm \gamma_5)/2$. 

For the antisymmetric tensor $\epsilon_{\mu\nu\rho\sigma}$ we take the sign 
convention $\epsilon_{0123}=1$.

\addcontentsline{toc}{chapter}{References}
\bibliographystyle{unsrt}
\bibliography{literatur1}

\begin{thebibliography}{10}

\bibitem{bpst}
A.A. Belavin, A.M. Polyakov, A.S. Schwartz, and Y.S. Tyupkin.
\newblock {\em Phys.~Lett.}, B 59:85, 1975.

\bibitem{tHooft76a}
G.~`t~Hooft.
\newblock {\em Phys.~Rev.~Lett.}, 37:8, 1976.

\bibitem{ILM_1}
E.~V. Shuryak.
\newblock {\em Nucl.~Phys.}, B 198:83.

\bibitem{ILM_2}
D.~I. Diakonov and V.~Y. Petrov.
\newblock {\em Nucl.~Phys.}, B 245:259, 1984.

\bibitem{Shuryak96}
T.~Sch\"afer and E.~Shuryak.
\newblock {\em Rev.~Mod.~Phys.}, 70:323, 1998.

\bibitem{BANCASH}
T.~Banks and A.~Casher.
\newblock {\em Nucl.~Phys.}, B 169:103, 1980.

\bibitem{DP}
D.~I. Diakonov and V.~Y. Petrov.
\newblock {\em Nucl.~Phys.}, B 272:457, 1986.

\bibitem{AGAPM95_01}
A.~Gonz\'alez-Arroyo and P.~Mart\'inez.
\newblock {\em Nucl.~Phys.}, B 459:737, 1996.

\bibitem{AGAYS}
A.~Gonz\'alez-Arroyo and Y.~Simonov.
\newblock {\em Nucl.~Phys.}, B 460:424, 1996.

\bibitem{COOLING}
M.~Teper.
\newblock {\em Phys.~Lett.}, B 171:81,86, 1986.

\bibitem{COOLING_2}
E.~M. Ilgenfritz, M.~L. Laursen, G.~Schierholz, M.~Muller-Preussker, and
  H.~Schiller.
\newblock {\em Nucl.~Phys.}, B 268:693, 1986.

\bibitem{SUSZCOOL}
P.~de~Forcrand, M.~Garc\'ia-P\'erez, and I.~O. Stamatescu.
\newblock {\em Nucl.~Phys.}, B 413:535, 1994.

\bibitem{APE}
M.~Albanese, F.~Costantini, G.~Fiorentini, F.~Flore, M.~Lombardo,
  R.~Tripiccione, and P.~Bacilieri.
\newblock {\em Phys.~Lett.}, B 192:163, 1987.

\bibitem{ALLES93}
B.~All\'es, M.~Campostrini, A.~DiGiacomo, Y.~G\"und\"uc, and E.~Vicari.
\newblock {\em Phys.~Rev.}, D 48:2284, 1993.

\bibitem{KIRCHNER97}
B.~All\'es, M.~D'Elia, A.~DiGiacomo, and R.~Kirchner.
\newblock {\em Phys.~Rev.}, D 58:114506, 1998.

\bibitem{hasenfratz98}
A.~Hasenfratz and C.~Nieter,
\newblock {\em  Phys.\ Lett.\ } {B 439}:366, 1998.

\bibitem{degrand98}
T.~DeGrand, A.~Hasenfratz and T.~G.~Kovacs,
\newblock {\em  Nucl.\ Phys.\ } {B 520}:301, 1998.


\bibitem{FERMION}
S.~Hands and M.~Teper.
\newblock {\em Nucl.~Phys.}, B 347:819, 1990.

\bibitem{Kaplan92a}
D.~B. Kaplan.
\newblock {\em Phys.~Lett.}, B 288:324, 1992.

\bibitem{Shamir}
Y.~Shamir.
\newblock {\em Nucl.~Phys.}, B 406:90, 1993.

\bibitem{ShamFur}
Y.~Shamir and V.~Furman.
\newblock {\em Nucl.~Phys.}, B 439:54, 1995.

\bibitem{Neuberger98a}
H.~Neuberger.
\newblock {\em Phys.~Lett.}, B 417:141, 1998.

\bibitem{Neuberger98b}
H.~Neuberger.
\newblock {\em Phys.~Lett.}, B 427:353, 1998.

\bibitem{DeGrand01}
T.~DeGrand and A.~Hasenfratz.
\newblock {\em Phys.~Rev.}, D 64:034512, 2001.

\bibitem{DeGrand02}
T.~DeGrand and A.~Hasenfratz.
\newblock {\em Phys.~Rev.}, D 65:014503, 2002.

\bibitem{Gattringer01}
C.~Gattringer, M.~G\"ockeler, P.~Rakow, S.~Schaefer, and A.~Sch\"afer.
\newblock {\em Nucl.~Phys.}, B 618:205, 2001.

\bibitem{Edwards02}
R.~Edwards and U.~Heller.
\newblock {\em Phys.~Rev.}, D 65:014505, 2002.

\bibitem{Blum02}
T.~Blum, N.~Christ, C.~Cristian, C.~Dawson, X.~Liao, G.~Liu, R.~Mawhinney,
  L.~Wu, and Y.~Zhestkov.
\newblock {\em Phys.~Rev.}, D 65:014504, 2002.

\bibitem{Horvath02b}
I.~Horv\'ath, S.~Dong, T.~Draper, N.~Isgur, F.~Lee, K.~Liu, J.~McCune,
  H.~Thacker, and J.~Zhang.
\newblock {\em Phys.~Rev.}, D 66:034501, 2002.

\bibitem{Teper02}
N.~Cundy, M.~Teper, and U.~Wenger.
\newblock {\em Phys.~Rev.}, D 66:094505, 2002.

\bibitem{Horvath03}
I.~Horvath {\it et al.}.
\newblock{Phys.\ Rev.\ } D  68:114505, 2003.

\bibitem{Horvath04}
I.~Horvath.
\newblock{Nucl.\ Phys. } B 710:464, 2005; Erratum-ibid.\ B  714:175, 2005.

\bibitem{Horvath05}
I.~Horvath {\it et al.}.
\newblock{Phys.\ Lett.\ } B 612:21, 2005. 




\bibitem{Hip02}
I.~Hip, T.~Lippert, H.~Neff, K.~Schilling, and W.~Schroers.
\newblock {\em Phys.~Rev.}, D 65:014506, 2002.

\bibitem{Gattringer02}
C.~Gattringer.
\newblock {\em Phys.~Rev.~Lett.}, 88:221601, 2002.

\bibitem{Horvath02c}
I.~Horv\'ath, S.~Dong, T.~Draper, F.~Lee, K.~Liu, H.~Thacker, and J.~Zhang.
\newblock {\em Phys.~Rev.}, D 67:011501, 2003.

\bibitem{Horvath02a}
I.~Horv\'ath, N.~Isgur, J.~McCune, and H.~B. Thacker.
\newblock {\em Phys.~Rev.}, D 65:014502, 2002.

\bibitem{THOOFT}
G.~`t~Hooft.
\newblock {\em Comm.~Math.~Phys.}, 81:267, 1981.

\bibitem{BAAL82}
P.~van Baal.
\newblock {\em Comm.~Math.~Phys.}, 85:529, 1982.

\bibitem{DK}
 S. Donaldson and P. Kronheimer The geometry of four-manifolds, 
\newblock {\em The geometry of four-manifolds}, Oxford UP,
1990.




\bibitem{ZEROMODES}
S.~Chada, A.~D`Adda, P.~DiVecchia, and F.~Nicodemi.
\newblock {\em Phys.~Lett.}, 67B:103, 1977.

\bibitem{ZEROMODES2}
R.~Jackiw and C.~Rebbi.
\newblock {\em Phys.~Rev.}, D 16:1052, 1977.

\bibitem{ADHM}
M.F. Atiyah, N.J. Hitchin, V.G. Drinfeld, and Y.~Manin.
\newblock {\em Phys.~Lett.}, A 65:185, 1978.

\bibitem{SIMMA}
T.~Kalkreuter and H.~Simma.
\newblock {\em Comput.~Phys.~Commun.}, 93:33, 1996.

\bibitem{ARNOLDI}
D.C. Sorensen.
\newblock {\em SIAM}, 13(1):357, 1992. The software is publicly available at
  http//:www.caam.rice.edu/software/ARPACK/.

\bibitem{QFIELDTHEOR}
P.~DiVecchia, K.~Fabricius, G.C. Rossi, and G.~Veneziano.
\newblock {\em Nucl.~Phys.}, B 192:392, 1981.

\bibitem{HEATBATH}
M.~Creutz.
\newblock {\em Phys.~Rev.}, D 21:2308, 1980.

\bibitem{LUSCHER82}
M.~L\"uscher.
\newblock {\em Nucl.~Phys.}, B 200:61, 1982.

\bibitem{PUGHTEPER_1}
D.~J.~R. Pugh and M.~Teper.
\newblock {\em Phys.~Lett.}, B 218:326, 1989.

\bibitem{PUGHTEPER_2}
D.~J.~R. Pugh and M.~Teper.
\newblock {\em Phys.~Lett.}, B 224:159, 1989.

\bibitem{Gattringer01_2}
C.~Gattringer, M.~G\"ockeler, P.~Rakow, S.~Schaefer, and A.~Sch\"afer.
\newblock {\em Nucl.~Phys.}, B 617:101, 2001.



\bibitem{Gattringer03}
C.~Gattringer and S.~Schaefer.
\newblock {Nucl.\ Phys.\ } B 654:30, 2003.

\bibitem{Gattringer04}
C.~Gattringer and R.~Pullirsch.
\newblock {Phys.\ Rev.\ } D {69}:094510,  2004.



\end{thebibliography}

\end{document}